\documentclass[10pt,journal,compsoc]{IEEEtran}

\ifCLASSOPTIONcompsoc
\usepackage[nocompress]{cite}
\else
\usepackage{cite}
\fi

\usepackage{subfigure}
\usepackage{wrapfig}
\usepackage{amsmath}
\usepackage{adjustbox}
\usepackage{amssymb}
\usepackage{subfigure}

\usepackage{multirow}
\usepackage{multicol}
\usepackage{graphicx}
\usepackage{caption}
\usepackage{makecell}
\usepackage{xcolor}
\usepackage[normalem]{ulem}
\usepackage{verbatim}
\usepackage{enumitem}

\newcommand {\bC}{\mathbf{C}}
\usepackage[ruled]{algorithm2e} 

\SetAlFnt{\small}
\SetAlCapFnt{\small}
\SetAlCapNameFnt{\small}
\SetAlCapHSkip{0pt}

\usepackage{color}
\definecolor{blue}{rgb}{0,0,0.8}
\definecolor{green}{rgb}{0,0.6,0}
\definecolor{red}{rgb}{0.7,0,0}

\begin{document}
\title{Visual Guidance for User Placement in Avatar-Mediated Telepresence between Dissimilar Spaces}

\author{Dongseok Yang,
        Jiho Kang,
        Taehei Kim, and
        Sung-Hee Lee% <-this % stops a space
\IEEEcompsocitemizethanks{\IEEEcompsocthanksitem Dongseok Yang, Jiho Kang, Taehei Kim, and Sung-Hee Lee are with Korea Advanced Institute of Science and Technology (KAIST).\protect\\
E-mail: \{dsyang, jhkang0408, hayleyy321, sunghee.lee\}@kaist.ac.kr
}% <-this % stops an unwanted space
}

\IEEEtitleabstractindextext{%
\setcounter{figure}{0} 
\begin{center}
    \centering
    \includegraphics[width=\linewidth]{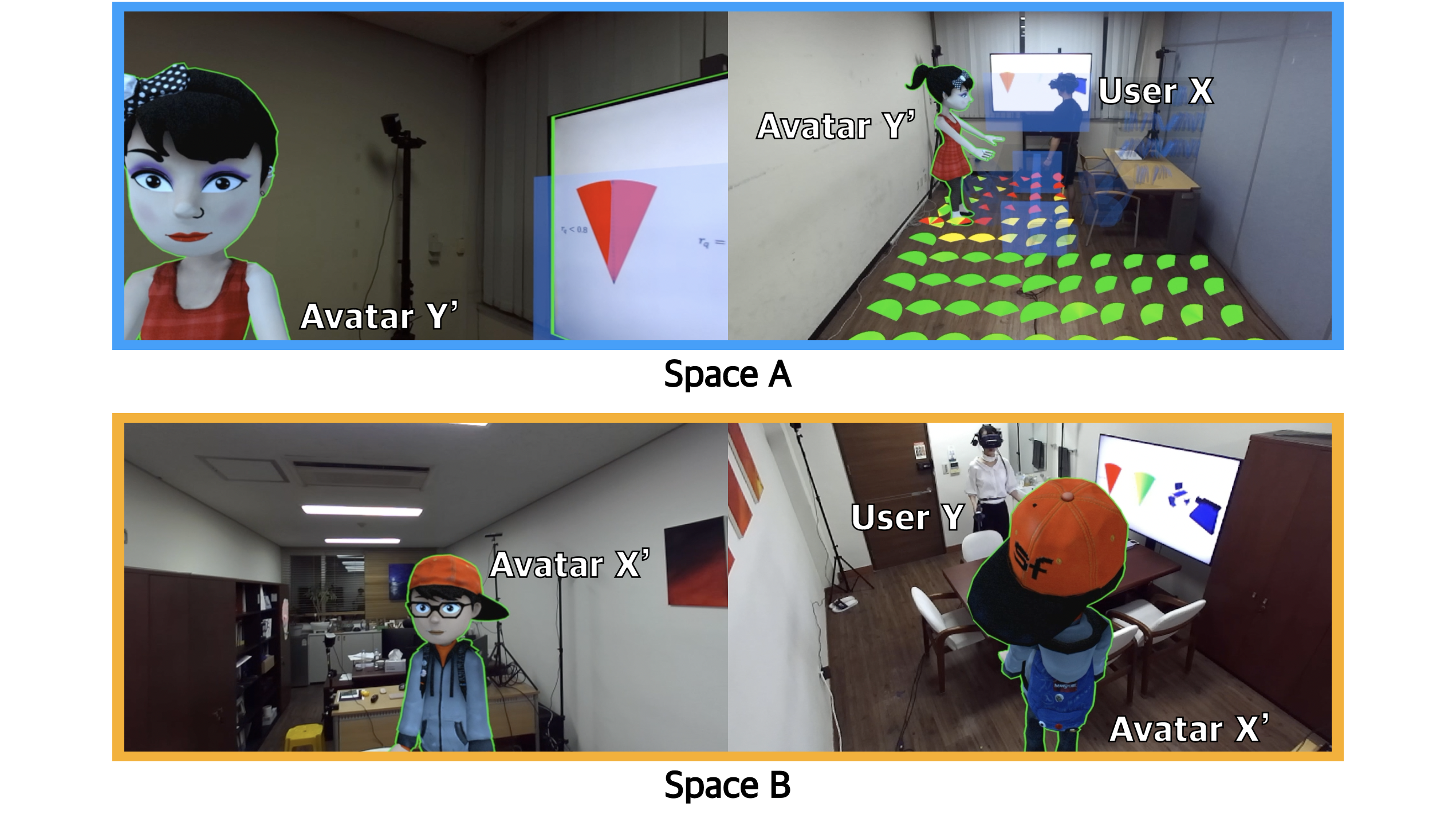}
    \captionof{figure}{User egocentric (left) and room perspective (right) views of space A (user X) and space B (user Y) in our MR telepresence system. Virtual avatars X' and Y' appear in remote spaces (space B for X', space A for Y') to represent user X and Y, respectively. When user X selects interaction targets (TV and Y' in this example) to interact with, our system provides visual guidance, color-coded sectors on the floor, and transparent 3D models of the remote space, to assist X in selecting his placement that will allow for his avatar X' to be appropriately placed to interact with the remote corresponding targets (TV and user Y in space B). After X arrives at his selected placement, our system places avatar X’ at an optimal location that best corresponds to user X's placement in space A, allowing bidirectional interaction between X and Y through their avatars.}
    \label{fig:teaser} 
\end{center}

\begin{abstract}
Rapid advances in technology gradually realize immersive mixed-reality (MR) telepresence between distant spaces. This paper presents a novel visual guidance system for avatar-mediated telepresence, directing users to optimal placements that facilitate the clear transfer of gaze and pointing contexts through remote avatars in dissimilar spaces, where the spatial relationship between the remote avatar and the interaction targets may differ from that of the local user. Representing the spatial relationship between the user/avatar and interaction targets with angle-based interaction features, we assign recommendation scores of sampled local placements as their maximum feature similarity with remote placements. These scores are visualized as color-coded 2D sectors to inform the users of better placements for interaction with selected targets. In addition, virtual objects of the remote space are overlapped with the local space for the user to better understand the recommendations. We examine whether the proposed score measure agrees with the actual user perception of the partner's interaction context and find a score threshold for recommendation through user experiments in virtual reality (VR). A subsequent user study in VR investigates the effectiveness and perceptual overload of different combinations of visualizations. Finally, we conduct a user study in an MR telepresence scenario to evaluate the effectiveness of our method in real-world applications.
\end{abstract}

}

\maketitle
\IEEEdisplaynontitleabstractindextext
\IEEEpeerreviewmaketitle

\IEEEraisesectionheading{\section{Introduction}\label{sec:introduction}}

\begin{figure}[t]
\includegraphics[width=\linewidth]{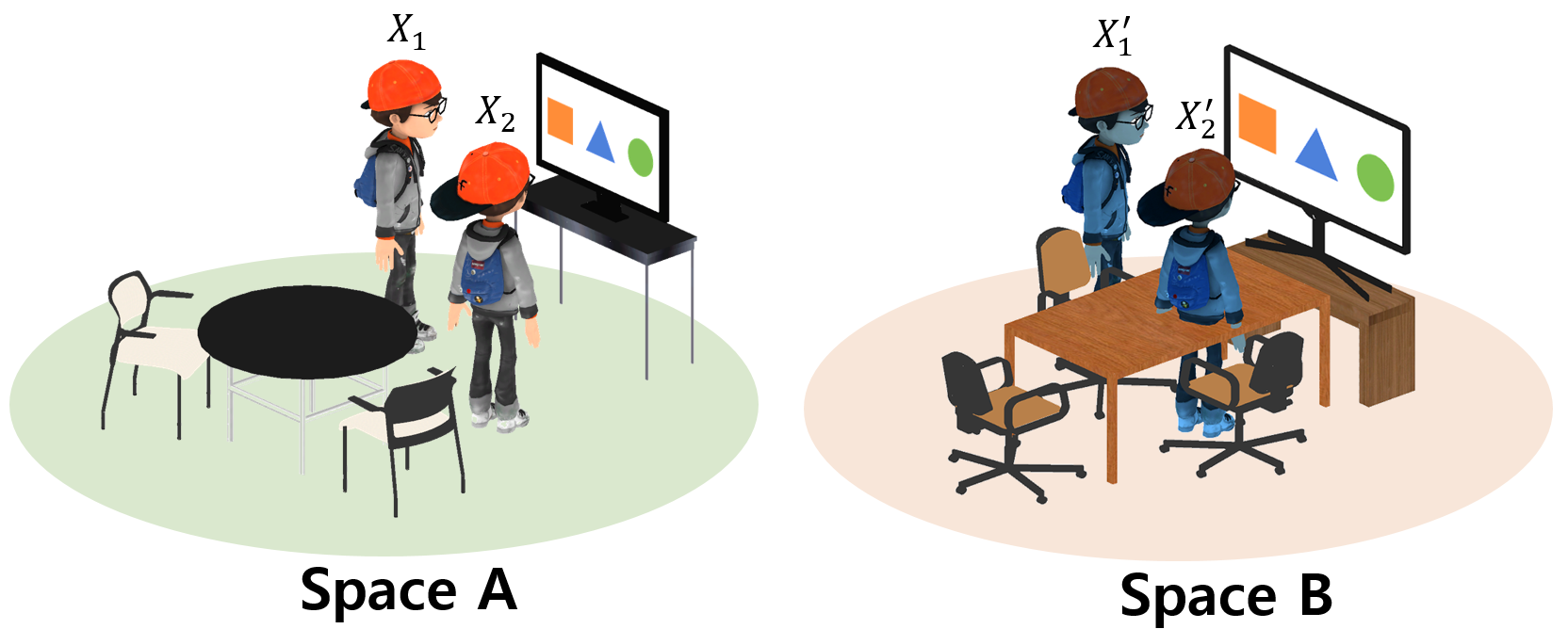}
\vspace*{-5mm}
\caption{ An example placement problem in MR telepresence between dissimilar spaces. In space A, both placements $X_1$ and $X_2$ suitably accommodate the interaction between the user and the target object (screen). However, their corresponding avatar placements that have the identical spatial relation between the avatar and the target object show that $X_2'$ is inappropriate due to the collision with a table, making $X_1$ a better placement than $X_2$. However, the users cannot predict the quality of the avatar placement that their placement will bring without additional information. This limitation emphasizes the need for supplementary guidance to enhance users' understanding and decision-making processes concerning avatar placement in MR telepresence scenarios.
}
\label{fig:problemdef}
\end{figure}

To realize seamless interaction between people in remote spaces, researchers have developed various telepresence approaches based on video, virtual reality (VR), and mixed reality (MR) \cite{ens2019revisiting}.
Among them, avatar-mediated MR telepresence has a prominent advantage that allows users to interact with physical objects in the local space while communicating with remote space users through virtual avatars.

As two physical spaces involved in telepresence may vary significantly in their shape and layout, a remote avatar should move adaptively to match the remote space's shape and objects in order to correctly convey the meaning of its user's motion taken in the local space. 
However, creating adaptive motions for avatars is a highly challenging task. Firstly, accurate inference of the meaning of a user's motion is difficult only with observable information, such as video data, unless the user confirms their intention. Given information on the user's motion semantics, the avatar needs to be placed and animated to preserve the meaning with respect to the remote space; the diversity of size, shape, and object layout of real spaces makes the problem more complex.

A straightforward way to avoid this challenge would be to restrict the user to be placed where a direct copy of the user's placement and motion into the avatar allows seamless interaction between remote users. For this, researchers have developed methods to find empty areas that can be shared by two spaces (e.g., \cite{keshavarzi2020optimization}). This approach, however, excludes areas occupied by objects from the telepresence area, reducing the size of available space for interaction. An approach to alleviate this limitation is to allow the user to utilize the entire space and place their remote avatar in the location that best preserves the meaning of the user's placement (e.g., \cite{yoon2020placement}). Our study follows this approach; Specifically, we put emphasis on preserving the local user's interaction context for avatar placement. Here, the \emph{interaction context} refers to a user's action on the target objects, dubbed \emph{interaction targets}, that the user is interacting with.

In this paper, we propose a novel practical approach to enhance the quality of avatar placement, which is to recommend good placements (position and orientation) for the local user that will result in high-quality avatar placements for preserving the user's interaction context. 

Our method is motivated by the fact that for each placement of a user in a given space, the quality of the corresponding placement of the avatar may be different.
For example, in Figure \ref{fig:problemdef}, user placements $X_1$ and $X_2$ are both appropriate for interacting with the screen. However, $X_1$ allows avatar placement $X_1'$ that exactly maintain the spatial relationship between the user (or avatar) and the screen while $X_2$ cannot. Therefore, $X_1$ is more advantageous than $X_2$ in terms of maintaining the spatial relationship. Since the user cannot see the remote space where the avatar is present, it is difficult to predict the quality of the avatar placement that their placement will bring without additional information. To solve this, our method visualizes and presents the quality of avatar placement to the user for each candidate user placement.

Figure \ref{fig:teaser} shows the screenshots of our prototype MR telepresence system. When a user selects the desired interaction targets, our method computes the quality of avatar placement, called \emph{recommendation score}, of sampled local placements and visualizes the scores with color-coded sectors on the floor to guide users to select an appropriate destination to interact with the desired targets (Fig. \ref{fig:teaser}, left). Our method includes an additional presentation of transparent 3D models of remote space. After a user confirms their arrival at a desired placement with an input device, our method places the remote avatar at the corresponding remote placement (male avatar in Fig. \ref{fig:teaser}, right).

To the best of our knowledge, this study is the first to provide visual guidance for better remote avatar placement in a bidirectional MR telepresence environment. Our approach is thoroughly validated through two user studies in a VR system simulating telepresence situations. Furthermore, we conduct a user study in an MR telepresence scenario to validate the effectiveness of visual guidance in practice.

Among various scenarios, we focus on a remote conference of two users in public meeting rooms or private rooms. We assume one-to-one preregistered correspondence between interaction target objects in both spaces (e.g., screens in two spaces correspond to each other). A virtual avatar mimics the motion of its corresponding user (synchronous avatar) and only its 2D position and orientation are controlled by our method. The range of interactions is narrowed to gaze and pointing gestures, which are primary non-verbal communications during a conference.

In summary, the main contributions of this paper are as follows:

\begin{itemize}
\item Visualization methods to guide a local user to placements whose avatar at the corresponding remote placements can well preserve the interaction context of the user, thereby enabling the remote user to correctly understand the local user's interaction context through the avatar's synchronous motion.

\item User experiments in VR and quantitative studies to validate that the proposed score measure is consistent with user perception of interaction context preservation, and visual guidance successfully drives users to select recommended placements.

\item A user experiment in MR telepresence environment and analysis to assess the effectiveness of our visual guidance and avatar placement method in the target application scenario.
\end{itemize}
\section{Related Work}

This section discusses previous studies on remote telepresence systems and visualizations in MR, which are the main focuses of this paper.

\subsection{Telepresence Systems}

Early telepresence research developed methods for visualizing remote users or surrounding objects in the local space using projection, display, and 3D capture technologies. Maimone et al. proposed a proof-of-concept telepresence system that realized real-time 3D scene capture and head-tracked stereo view using multiple RGB-d cameras \cite{Maimone2011Encumbrance} and elaborated the system with a customized see-through head-worn display and a projector that merges the remote user's visual information into the user's local environment \cite{maimone2013general}.
Steed et al. \cite{Steed2012Beaming} introduced a system that invites remote users to the local physical destination. Remote visitors wear motion tracking suits and their movements are transmitted to animate avatars in the local space. Surround cameras in the local system capture image of the local destination and send it back to the remote space to be rendered in visitor's HMDs.
Beck et al. \cite{beck2013immersive} reconstructed virtual images of two remote groups of people from depth images to a shared virtual world with a projection-based method. A novel concept of the cylindrical display was proposed by Pan et al. \cite{pan2014gaze}, allowing the users to correctly perceive remote user's gaze direction and eye contact; such non-verbal cues play an important role in face-to-face interactions but are frequently lost in 2D planar displays.
As the real-time spatial capture and reconstruction became possible with a single HMD, researchers developed an end-to-end MR telepresence system \cite{orts2016holoportation} that reconstructs volumetric meshes of remote objects and users in the local space. The straightforward capture-and-reconstruct approach does not consider dissimilarity between remote spaces, but the spatial dissimilarity makes it challenging to convey the user's intention correctly. Addressing this problem, researchers proposed methods to define a valid area for interaction. Lehment et al. \cite{lehment2014creating} optimized the alignment between two remote rooms to form a consensus space with the maximum common features. 

Another area of research focuses on Avatar-mediated telepresence which enables user embodiment \cite{roth2019technologies}, in contrast to RGB-d replication-based systems. In Avatar-mediated telepresence, a crucial aspect is determining the appropriate placement of the remote avatar to effectively convey the user's context to partners in the remote space. Room2Room \cite{pejsa2016room2room} introduced a heuristic scheme to determine the ideal placement of a remote avatar that corresponds to the local user seating or standing. Yoon et al. \cite{yoon2020placement} developed a deep neural network trained with placement data obtained from a user experiment to compute the remote placement that best preserves the geometric relation between the local user and their surrounding objects. These studies still have limitations since they do not consider the follow-up interpersonal and human-object interactions. On the other hand, our visual guidance recommends local placements according to the degree of preserving gaze and pointing context.

Several recent studies focused on various techniques to support MR-based remote collaboration. Piumsomboon et al. \cite{piumsomboon2018mini} presented Mini-Me, a size-adjustable avatar that transforms its scale and orientation to adapt to the remote user's field of view (FOV) while maintaining the local user's gaze and gesture. Kumaravel et al. \cite{thoravi2019loki} proposed Loki, a bi-directional MR telepresence system with user interfaces of VR/AR view switching, 2D video, and hologlyph. Young et al. \cite{young2019immersive} provided the users with a panoramic representation of surroundings and rendered the current FoV and hands of the remote user. While these methods assume the use of a shared space or interaction in only one space, our system allows users to simultaneously utilize the two spaces. This is particularly important considering real-world spaces, which typically lack a dedicated XR space without furniture. The concept of Partially Blended Realities by Grønbæk et al. \cite{gronbaek2023partially} addresses a problem akin to ours. While their approach involves selective alignment of two remote spaces according to a single target object, our method supports multiple interaction targets and recommends suitable user placements for interaction.

\subsection{Visualizations in Mixed Reality}

Giving visual cues for effective communication in MR has been researched since the emergence of the technology. Early works focused on improving user performance in a specific task by drawing user attention with simple visual cues and annotations \cite{Rusch2013Drivercue, marner2013annotations, volmer2018proceduralcue, Reyes2016MachineryAR, irlitti2019conveying, Nylin2020SoftCue} or even with a digital copy of physical objects \cite{yu2022duplicated}. These studies increased the user's competence and proved the potential of MR visualizations. However, these systems do not explicitly consider remote collaboration in MR. Moreover, as Ishii et al. \cite{Ishii1994IterativeDO} pointed out, such 2D annotations are not suitable for MR environments where users have free access to both the physical and the virtual contents.

Since gaze and pointing are essential to understanding user interaction context and thus critical for immersive MR telepresence collaboration, many works focused on sharing the information by providing visual cues. Gupta et al. \cite{gupta2016gazetracking} studied visualization of the remote helper's gaze and pointing in the local worker's live HMD view. They compared different conditions of visualizing the remote user's pointing and gazing to the local user building given structures with LEGO blocks. The experiment showed that providing both cues helped users understand each other, significantly increasing the sense of co-presence. Piumsomboon et al. \cite{Piumsomboon2017Enhancements} investigated the embodiment of the avatar representation with the head and hands. In addition, they found that visualizing the virtual boundary of FoV enhances communication during MR collaboration. As follow-up research on the effects of visual cues, they compared combinations of visual cues, including FOV frustum, eye-gaze ray, and head-gaze ray \cite{Piumsomboon2019threetypecue}. They designed symmetric searching and asymmetric placing tasks of virtual blocks in a shared space of one physical (AR) and one virtual (VR) setup. The mixture of FOV frustum and the head-gaze ray was found to bring the highest task performance and preference. Bai et al. \cite{bai20203dpanorama} implemented a 3D panorama-based MR collaboration system and conducted a user study to investigate the effect of adding hand gesture cues on context understanding and co-presence. They experimentally proved that the combined cues of gaze and gesture deliver spatial actions significantly better than the gaze cue alone. Despite their observations on the effects of visualizations in MR collaboration, these studies suppose the MR system of relocating remote users to the local user's space.

Recent technical progress, such as spatial capture and real-time tracking, allowed researchers to introduce novel visualization methods for collaboration in a bidirectional MR environment. Several studies captured the remote space in real-time and reconstructed them as point clouds \cite{thoravi2019loki} or 3D meshes \cite{orts2016holoportation} in the local space. Chenechal et al. \cite{Chenechal2016Vishnu} introduced virtual arms of the remote expert as interactive guiding tools for the local user during collaborative work. Gurevich et al. \cite{Gurevich2015TeleAdvisor} proposed a hands-free remote AR projection system consisting of two cameras on the teleoperated robotic arm to stream the local worker's space; the remote helper sees the physical space for effective communication. Teo et al. \cite{Teo2019Combining} developed a hybrid system of 360 panorama video and reconstructed 3D scenes and compared the two methods through a user study. Participants reported that panorama view is better for figuring out the partner's attention while the reconstructed scene is better for performing given tasks.

While previous studies dealt with visualizing remote space itself and user interfaces allowing easy annotation and manipulation, we focus on modeling processed information of recommendation scores of local placements and visualizing them to support clear communication between distant users.
\section{Method}

The purpose of our system is to achieve a clear transfer of the user's interaction context to the corresponding avatar in the remote space. Since our method is used before users take action, interaction targets (the target objects for gaze and pointing) are manually set by user input. Then, our system computes and visualizes the recommendation scores of local placements as shown in the left of Figure \ref{fig:teaser}. The visual guidance encourages the users to move to a better placement such that their avatars can be located to preserve the user's interaction context for the selected targets -- For each sampled local placement for visual guidance, our system computes the optimal corresponding placement (\textbf{OCP} for short), the corresponding remote placement that best preserves the interaction context of the local placement and the associated recommendation score. After the user arrives at a desired placement, our method places the remote avatar at its OCP; the right images of Figure \ref{fig:teaser} show the remote avatar placed at the OCP. Our method deals with both interpersonal and human-object relations; for our experiments, the screen and the other user's avatar can be chosen as interaction targets. 

\subsection{Interaction Feature}
\label{subsec:interaction_feature}

We start with formulating the feature of the placement with respect to the interaction context, which will be used as a measure for placing the avatar. 

Recent studies on Collaborative Virtual Environments (CVEs) \cite{mayer2020improving, sousa2019warping} found that the observer's viewpoint from the target affects the accuracy of understanding of the other users' gestures in a shared virtual space. Different from those works in CVEs, our system is designed for MR collaboration between two distant real spaces. However, we still share the goal of increasing the accuracy of interpreting gaze and gesture when the user can only observe the other user's avatar. To achieve this, we hypothesize that the spatial relationship between the user and interaction targets of the local space should be preserved as much as possible for the corresponding targets in the remote space.

For this, we first define the \emph{interaction feature} $\Phi$ that represents the spatial relation between a source object $s$ and a target object $t$ as follows:
\begin{equation}
\Phi = [\phi_{s \rightarrow t}^{R},\ \phi_{s \rightarrow t}^{L},\ \phi_{t \rightarrow s}^{R},\ \phi_{t \rightarrow s} ^{L}],\ -\pi < \phi \leq \pi,
\end{equation}
where $\phi_{s \rightarrow t}^{R}$ denotes the angle between the frontal direction of $s$ to the right end of $t$. Given the placements (2D position and orientation) of the source and target $q_s = (x_s, y_s, \theta_s)$, $q_t = (x_t, y_t, \theta_t)$, and their right/left endpoints $(x_s^R, y_s^R)$ / $(x_s^L, y_s^L)$, $(x_t^R, y_t^R)$ / $(x_t^L, y_t^L)$, the interaction feature is calculated as
\begingroup\makeatletter\def\f@size{8}\check@mathfonts
\begin{align*}
    \Phi = [&\textrm{atan2}(y_{t}^{R} - y_{s},\ x_{t}^{R} - x_{s}) - \theta_s, \
    \textrm{atan2}(y_{t}^{L} - y_{s},\ x_{t}^{L} - x_{s}) - \theta_s, \\
    &\textrm{atan2}(y_{s}^{R} - y_{t},\ x_{s}^{R} - x_{t}) - \theta_t, \
    \textrm{atan2}(y_{s}^{L} - y_{t},\ x_{s,}^{L} - x_{t}) - \theta_t].
\end{align*}
\endgroup

\begin{figure}[t]
\centering
\includegraphics[width=0.95\linewidth]{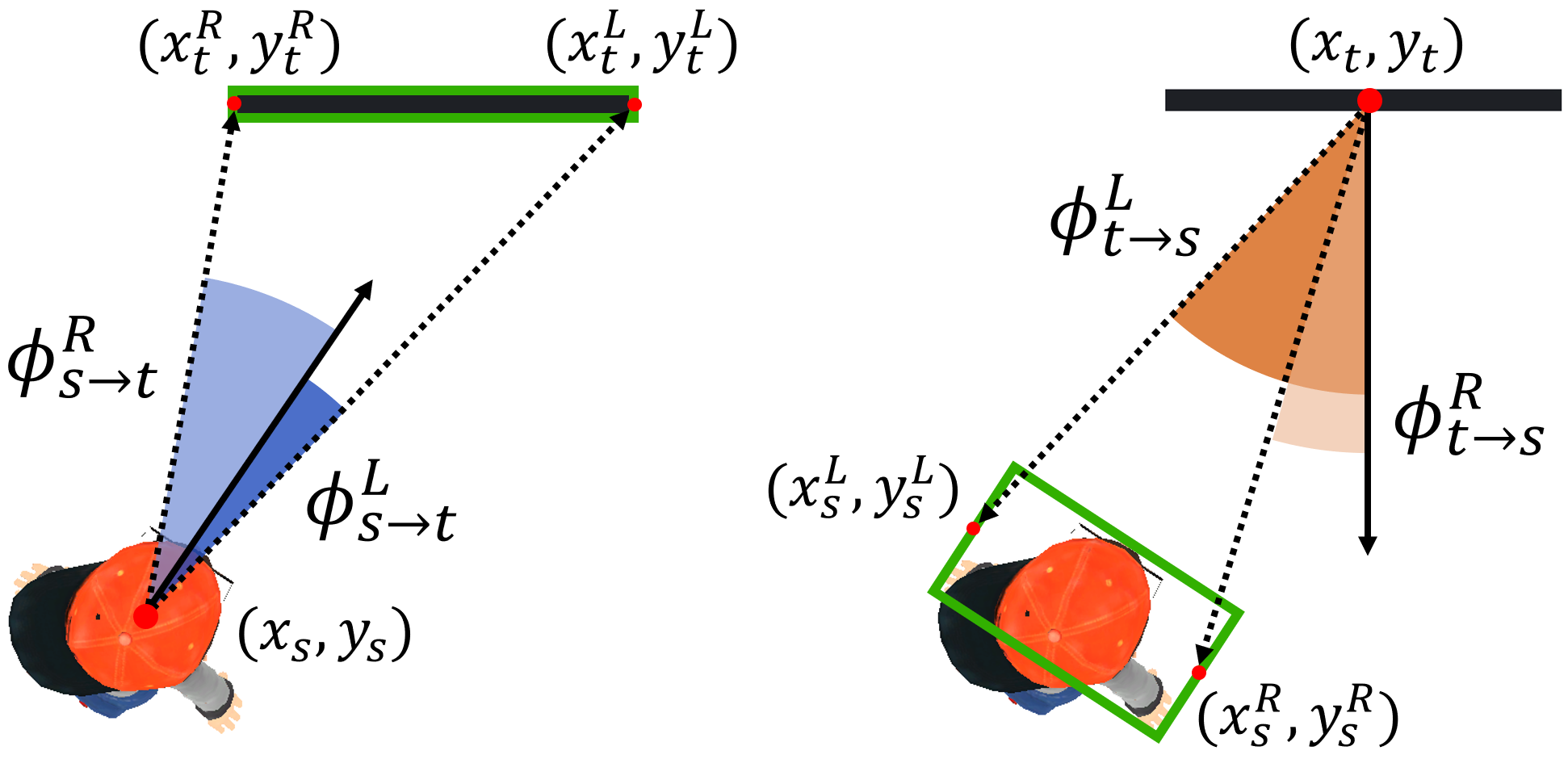}
\caption{Four angles that build the interaction feature between a source object (a user, $s$) and a target object (a screen, $t$). Green rectangles on objects represent the bounding boxes for defining endpoints.}
\label{fig:angle difference representation}
\end{figure}

Figure \ref{fig:angle difference representation} shows an example interaction feature between a user and a screen. Note that we also consider the angles from $t$ to $s$ to preserve the direction from $t$ to $s$. 

\begin{figure}[t]
     \centering
     
     \subfigure[]
     {
         \includegraphics[width=2.0in]{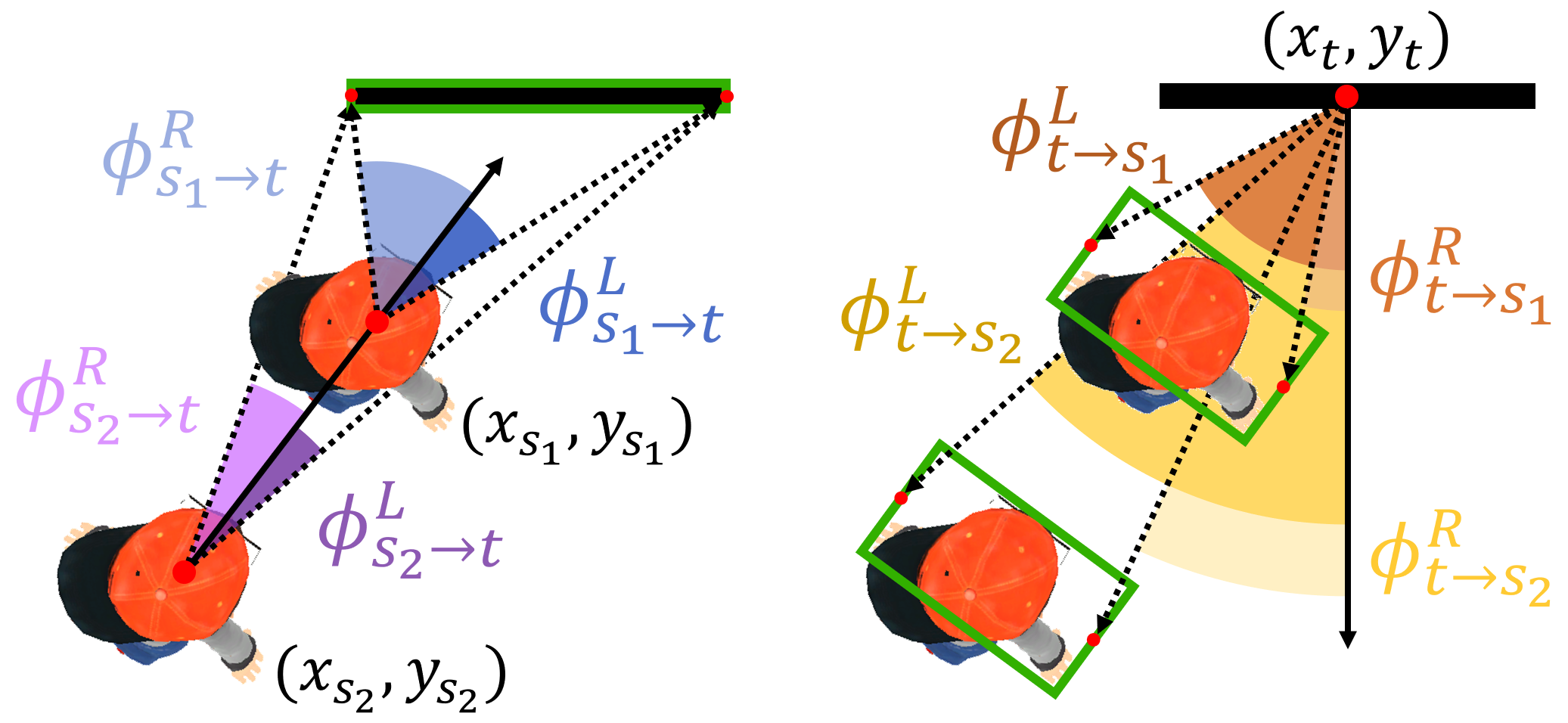}
         \label{fig:angle difference representation example1}
     }
     \subfigure[]
     {
         \includegraphics[width=1.3in]{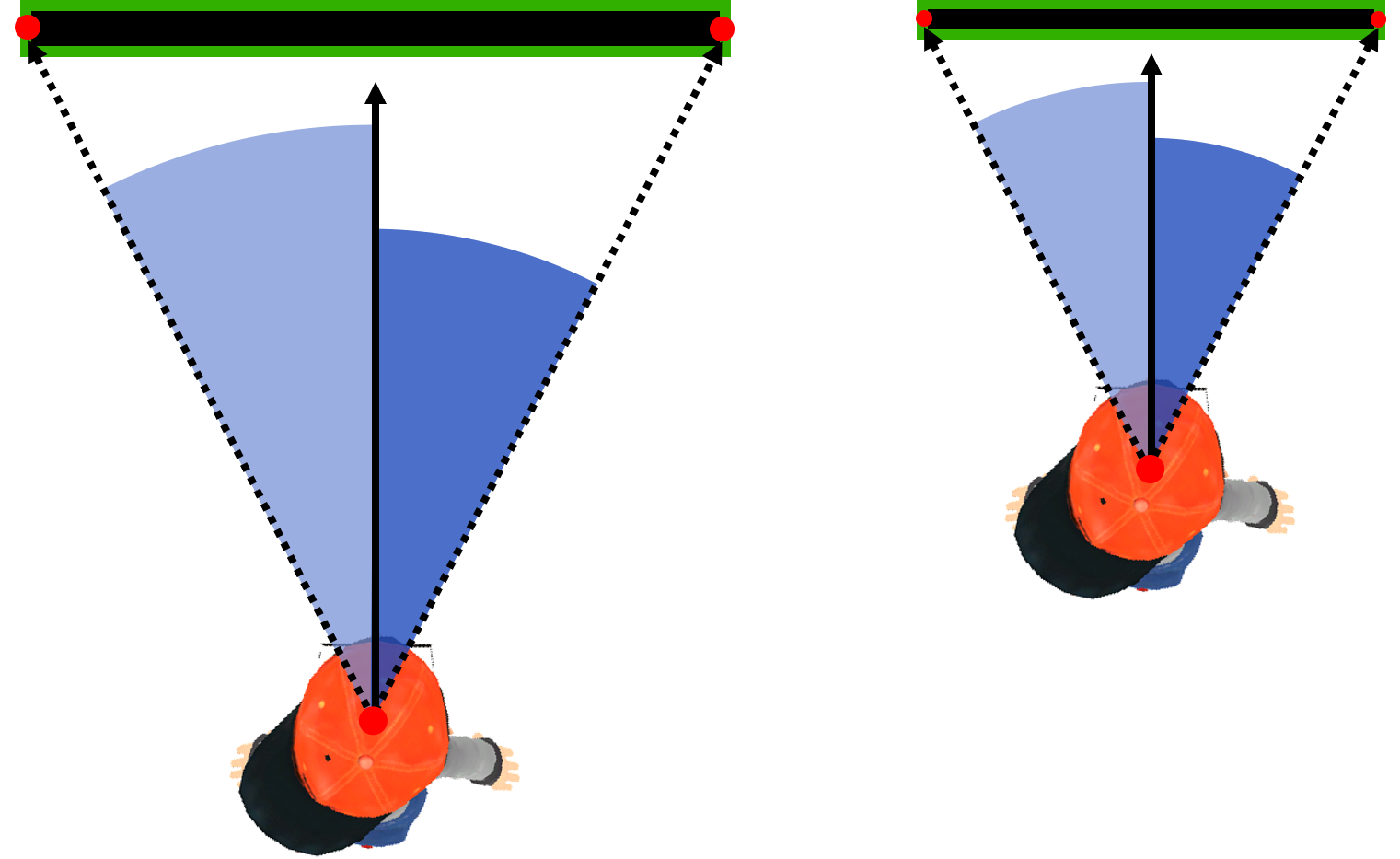}
         \label{fig:angle difference representation example2}
     }
    \caption{(a) Interactions features for two placements with the same direction but different distance from the target $(x_t,y_t)$. (b) Two placements with the same features $\phi_{s \rightarrow t}$ for targets with different sizes.}
    \label{fig:interaction feature examples}
\end{figure}

By using four angle attributes between two objects, the interaction feature implicitly accounts for the distance between two objects, not just the directions. For example, in Figure \ref{fig:interaction feature examples}(a), $s_1$ and $s_2$ are in the same direction from $t$, but their interaction features $\Phi_1$ and $\Phi_2$ are different; {$\phi_{s_1 \rightarrow t}^{R}$, $\phi_{s_1 \rightarrow t}^{L}$, and $\phi_{t \rightarrow s_1}^{L}$ are larger than $\phi_{s_2 \rightarrow t}^{R}$, $\phi_{s_2 \rightarrow t}^{L}$, and $\phi_{t \rightarrow s_2}^{L}$, respectively, and $\phi_{t \rightarrow s_1}^{R}$ is smaller than $\phi_{t \rightarrow s_2}^{R}$.}
Since we use only angles to define the interaction feature without using distance, preserving the interaction features for targets with different sizes in two distant spaces means that the interaction targets take the same position and size in the egocentric visions of the user and the avatar as shown in Figure \ref{fig:interaction feature examples}(b).

The feature vector of an interaction between the source and $n$ interaction targets is defined as a concatenation of the interaction features for each target, $\boldsymbol{\Phi} = [\Phi_i]_{i=1}^{n}$.

As the avatar mirrors the user motion, we hypothesize that sharing the same interaction feature between a local user and their remote avatar allows effective remote communication; a remote partner can fully understand the local user's intention on targets (interaction context) only by observing the behavior of local user's avatar. Therefore, we define the similarity between local (a user) and remote (their avatar) interaction features as a quantitative measure for the degree of interaction context preserved.

Given the local placement $q=(x, y, \theta)$ and remote placement $q'=(x', y', \theta')$, we define the interaction feature similarity as the Gaussian kernel distance: 

\begin{equation} \label{eq:similarity}
S(\boldsymbol{\Phi}_{q},\ \boldsymbol{\Phi}_{q'}) = e^{-2 ||\boldsymbol{\Phi}_{q} - \boldsymbol{\Phi}_{q'}||^2}.
\end{equation}

\begin{figure}[t]
\centering
\includegraphics[width=0.95\linewidth]{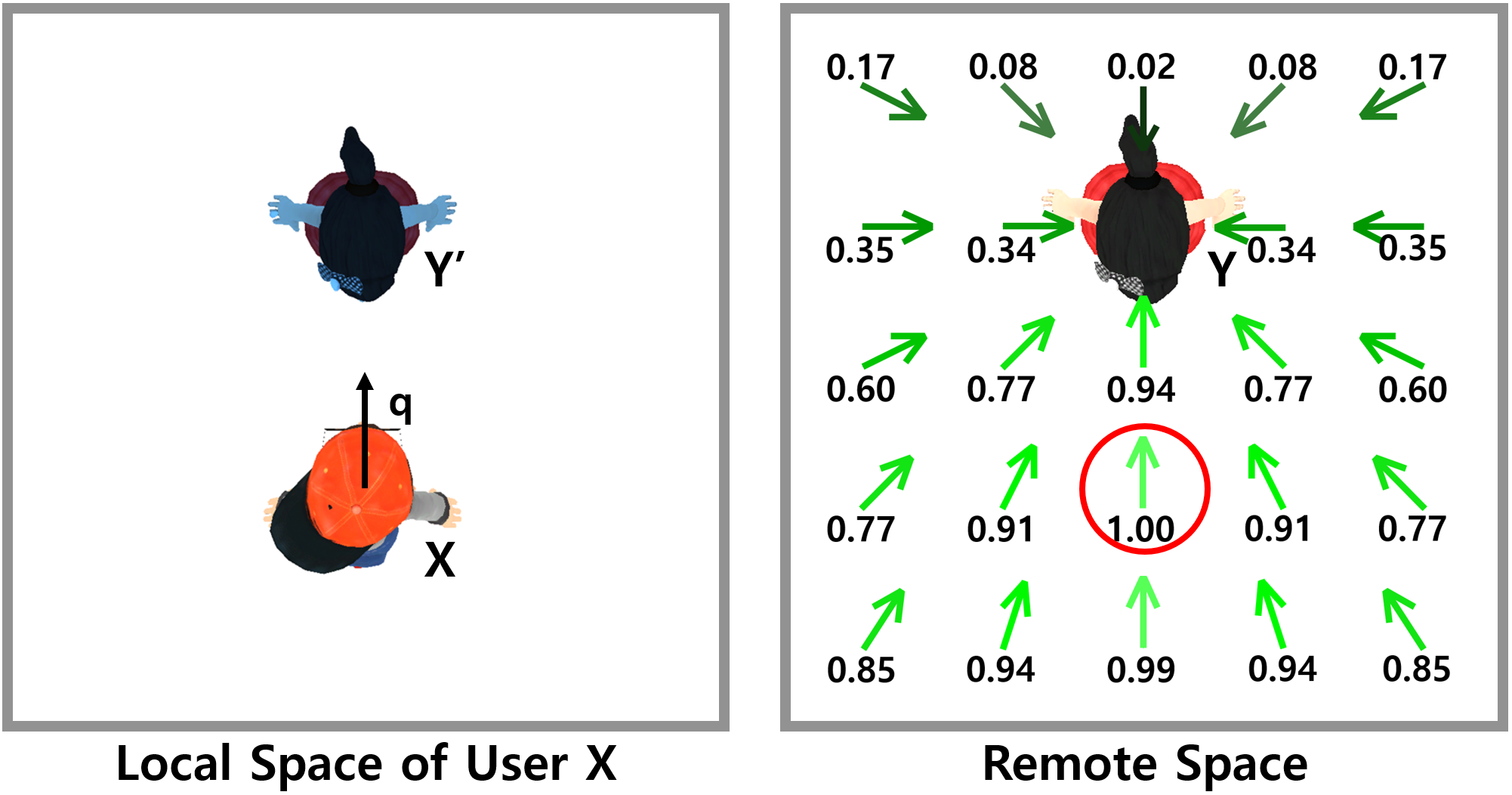}
\caption{Feature similarities of sample placements of the avatar X', shown as arrows. The interaction target is set as Y for X', and Y' for X. The green color corresponds to a similarity of 1.00, and the smaller the similarity, the closer the color is to black.} 
\label{fig:similarity}
\end{figure}

Figure \ref{fig:similarity} shows feature similarities of sampled placements of the local user's avatar X' in the remote space for a given X placement in the local space, considering the interpersonal interaction (X-Y' and X'-Y). The identical placement of the avatar in the remote space (marked with the red circle) has the highest similarity of 1.0 and the similarity gradually decreases as the placement deviates in angle or distance from the identity.
The validity of our interaction feature and the similarity measure is examined in Section \ref{sec:Exp1}.

\begin{figure*}[t]
\centering
\includegraphics[width=0.9\linewidth]{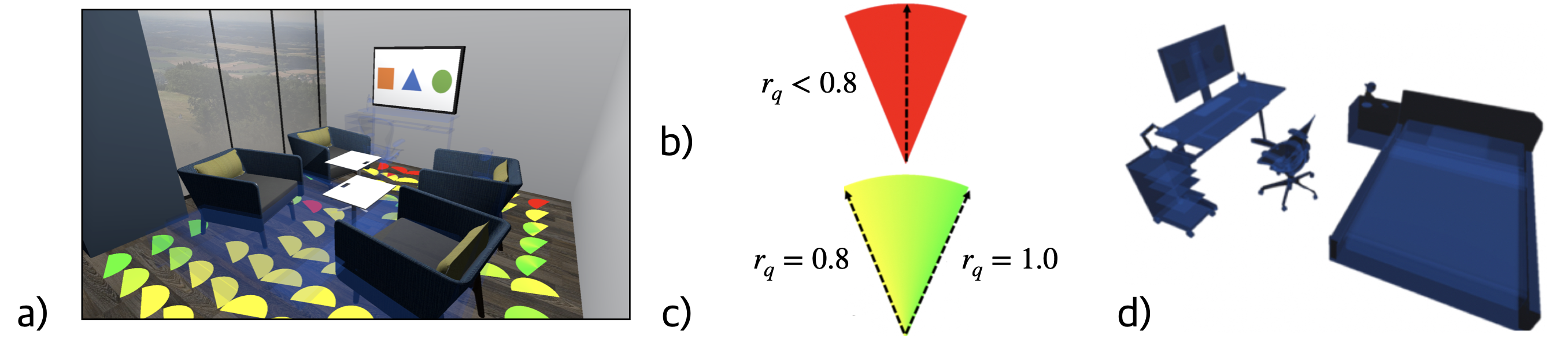}
\caption{Overview of our visual guidance. a) An example of visual guidance seen from a user's perspective. b) A red sector represents a placement with $r_q$ lower than the threshold ($0.8$). The central angle of each sector equals the interval of direction sampling ($\frac{\pi}{4}$ for experiments). c) For placement with $r_q$ higher than the threshold, we color the line according to $r_q$ (0.8 = yellow and 1.0 = green). d) Transparent copies of remote space objects are shown in the user space with respect to the coordinate frames of the paired interaction targets.}
\label{fig:visoverview}
\end{figure*}

\subsection{Recommendation Scores}
Due to the spatial discrepancy in MR telepresence, it is impossible to have an identical interaction feature for local and remote placements in many cases. Therefore, our system recommends local placements that have a certain degree of interaction context preservation, measured as the interaction feature similarity, or above. 

For a candidate placement $q$, we compute its recommendation score $r_q$ as the maximum interaction feature similarity obtainable from the remote space:
\begin{equation} \label{eq:r_q}
r_q = \max_{q' \in \widehat{Q}'} S(\boldsymbol{\Phi}_{q},\ \boldsymbol{\Phi}_{q'}),
\end{equation}
where $q'$ is a placement in the continuous space of feasible placements $\widehat{Q}'$ in the remote space. Here, feasible placement refers to a placement where the avatar can exist naturally, free from a collision with furniture or wall.

The remote placement $q'$ that has the highest feature similarity for given local placement $q$ is defined as the OCP of $q$, denoted as $q^*=(x^*, y^*,\theta^*)$.

\subsubsection{Finding Optimal Corresponding Placement}
To find the OCP $q^*$ as a feasible placement, we consider not only its interaction feature similarity but also its feasibility to accommodate the avatar. For this, we add a collision cost and a out-of-space cost to the objective function Eq. \eqref{eq:r_q}. \\

\noindent \textit{\textbf{Collision cost.}} The collision cost is designed to avoid collision between avatar and objects; the function is defined as a multivariate Gaussian function with mean $p'_{obj}$ at the center of an object and standard deviation determined by the object's width $w'_{obj}$, length $l'_{obj}$, and orientation $\theta'_{obj}$.

\begin{equation}
C_{col,\ q'} = e^{-\frac{1}{2}[p' - p'_{obj}]\Sigma^{-1}[p' - p'_{obj}]^T} 
\end{equation}
\[
\Sigma = \text{R}\Lambda\text{R}^{\text{T}}
\]
\[
\text{R} = \begin{bmatrix}
           \cos{(\theta_{obj}^{\prime})} & -\sin{(\theta_{obj}^{\prime})} \\
            \sin{(\theta_{obj}^{\prime})} & 
            \cos{(\theta_{obj}^{\prime})}
           \end{bmatrix}, \ 
\Lambda = \begin{bmatrix}
           (\frac{w_{obj}^\prime}{2})^2 & 0 \\
            0 & (\frac{l_{obj}^\prime}{2})^2
           \end{bmatrix}
\]
\newline

\noindent \textit{\textbf{Out-of-space cost.}} To limit the OCP inside the remote space during optimization, we design out-of-space cost defined as an exponential ``cliff'' function along the four borders of the space floor, in which $x'_c$ and $y'_c$ are the center coordinates of the space, and $w'$ and $l'$ are the space's width and length.

\begin{equation}
C_{out,\ q'} = 3 ( e^{\frac{2}{w'}\cdot(\lvert x'-x'_{c} \rvert - \frac{w'}{2})}+e^{\frac{2}{l'}\cdot(\lvert y'-y'_{c} \rvert - \frac{l'}{2})} )\\
\end{equation}
\

\noindent The OCP $q^*$ is obtained by the steepest descent algorithm that iteratively updates the remote placement $q'$ with gradient $\nabla \bC$ computed according to the condition of $q'$:

\[
q'_i \leftarrow q'_i + \gamma \nabla \bC
\]

\[
\label{gradient}
\nabla \bC =
\begin{cases}
-\frac{dC_{col, q'}}{dq'} & \text{if } q'\ \text{in collision}, \\
-\frac{dC_{out, q'}}{dq'} & \text{if } q'\ \text{out of space}, \\
\frac{dS}{dq'} & \text{otherwise}.
\end{cases}
\]
We determine that $q'$ is in collision if it intersects any predefined bounding box of an object and $q'$ is out of space if it lies outside the bounding box of the floor.

For our experiments, the step size $\gamma = [\gamma_{x'},\ \gamma_{y'},\ \gamma_{\theta'}]$ was set as $[0.1,\ 0.1,\ 1.0]$ and the maximum iteration count was set as $180$. After obtaining $q^*$, we set the feature similarity $S(\boldsymbol{\Phi}_{q},\ \boldsymbol{\Phi}_{q^*})$ as the recommendation score $r_q = S(\boldsymbol{\Phi}_{q},\ \boldsymbol{\Phi}_{q^*})$ of a candidate local space placement $q$. % \in Q^+$: $r_q $.

\subsubsection{Local Placement Sampling}
For computational feasibility, we discretize the local space into $n$ sample placements and compute their recommendation scores. The system first generates a set $Q$ of local placement samples by grid-sampling the space with distance and angle intervals ($0.33$m and $\frac{\pi}{4}$ for our experiments) to obtain $Q = \{q_i\}_{i=1}^{n},\ q_i = (p_i,\ \theta_i),\ p_i = (x_i,\ y_i)$. 
According to Strasburger's report on the horizontal span of human vision \cite{strasburger2020seven}, we set visibility constraint as a horizontal range of $\pm\frac{\pi}{2}$ from an egocentric viewpoint. Among the samples in $Q$, we include only $q_i$ in a candidate sample set $Q^+$ only if the source $q_i$ and target are mutually visible (i.e., all the absolute values of angle differences between the source angle and the angles of vectors from the source position to the target position are smaller than $\frac{\pi}{2}$). For every element $q_i \in Q^+$, we compute its OCP in the remote space and the corresponding recommendation score.

\subsection{Visual Guidance}

To visualize the scores to the user, we propose to use 2D colored sectors. In addition, 3D transparent models of the remote space objects are overlapped on the local space to provide a hint of the remote space layout. The main goal of the visual guidance is to influence the users to choose a better local placement so that the avatar placed at the OCP can effectively deliver the user's interaction context to the other user. In practice, our system provides visual guidance when interaction targets are specified by user input. Given the information, users can freely move to the placement of their choice for interaction and place their remote avatar at OCP. \\

\noindent \textit{\textbf{2D Sectors.}}
Figure \ref{fig:visoverview} b) and c) show images of 2D sectors in the top view. Each placement $q = (x, y, \theta) \in Q^{+}$ is represented as a 2D cone with a central angle set to the sampling interval ($\frac{\pi}{4}$ for experiments), located at $(x, y)$ with a direction of $\theta$. If the recommendation score $r_q$ is below a threshold ($0.8$), we color the entire cone red to inform the users that the placement is inappropriate for the interaction with the assigned targets; the threshold value is found through a user experiment as described in Sec. \ref{sec:Exp1}. Recommendation scores of $0.8$ and $1.0$ correspond to the yellow and green, and the colors of in-between scores are linearly interpolated in the HSV color space. Since we only sample eight orientations for each position, the scores of in-between directions are estimated by linearly interpolating scores of two adjacent samples. \\

\noindent \textit{\textbf{Overlapped Transparent Models.}}
Only with sectors visualizing the recommendation scores of placements can the users not understand why a specific placement is inappropriate for interaction. Therefore, to make our visual guidance more informative, we show transparent models of the remote space objects in the local space. Figure \ref{fig:visoverview} d) shows an image of transparent remote models. If the user selects local objects as interaction targets, 3D models of the remote corresponding objects are overlapped with respect to the coordinate frames of the primary target. For multiple interaction targets, we set a screen as the primary target. Visualizing remote objects allows users to check the identical placement in the remote space, and brings the additional benefit of helping them choose placements that give better visibility for the remote avatar on the remote interaction targets. \\

\noindent \textit{\textbf{Top View.}}
Some 2D sectors can be occluded by real furniture and objects, and thus the users may not see the recommendations for the entire placements. Such partial observation may restrict the users' movement only within visible areas. To prevent this bias, we allow the user to optionally see a top view of the local space with visual guidance; the top view can be enabled/disabled by user input.
\section{Evaluation}

To evaluate the proposed recommendation score and visual guidance, we conduct two experiments in VR. The first experiment aims to examine whether the proposed score is consistent with the users' actual perception of the degree of interaction context preserved during the remote telepresence. The second experiment evaluates whether our visual guidance successfully leads the users to the placements whose OCPs are appropriate for interacting with the selected targets. Furthermore, we validate the effectiveness of our method in a user experiment conducted in a target MR telepresence environment, specifically in a remote presentation scenario between distant spaces.

\begin{figure*}[h]
\centering
\includegraphics[width=0.9\linewidth]{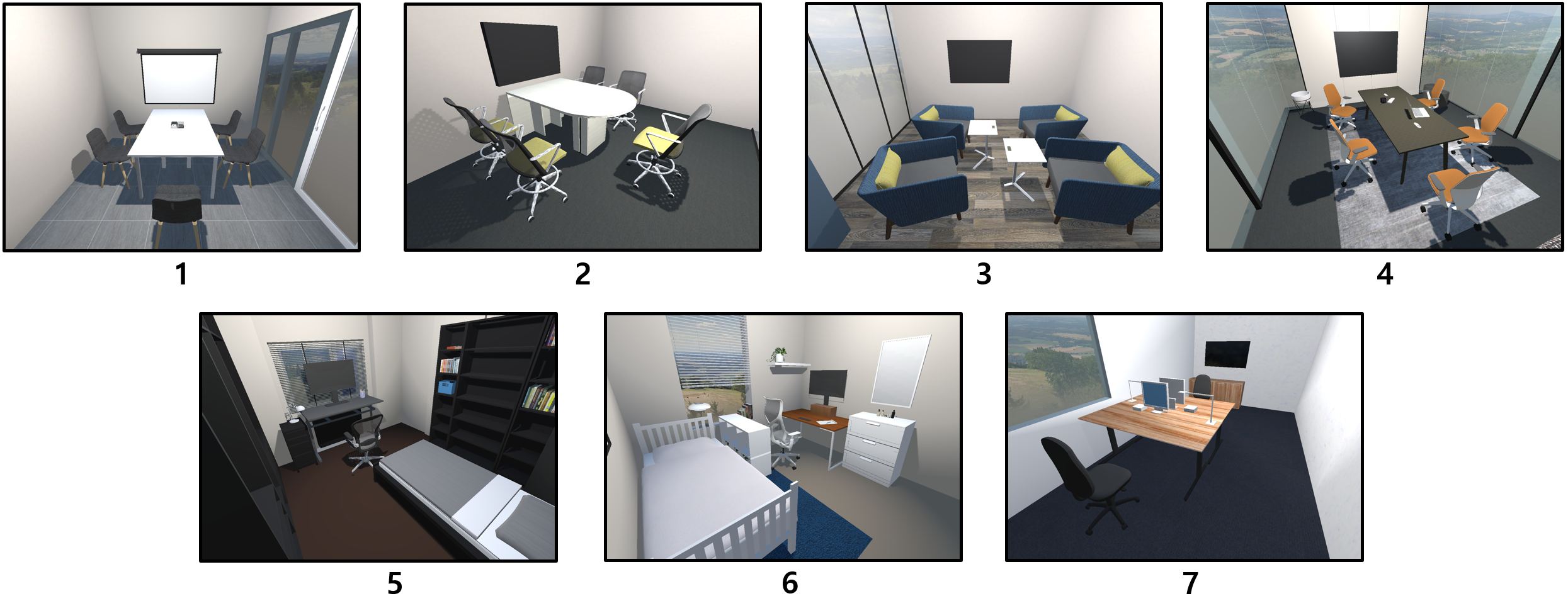}
\caption{Perspective view images of virtual space models. We select 4 conference rooms (Space 1-4) and 3 private rooms (Space 5-7) of different sizes and layouts to validate the proposed method for varying spatial discrepancy between two distant spaces. }
\label{fig:vrspaces}
\end{figure*}

\begin{table*}[h]
 \centering
 \begin{tabular*}{\textwidth}{l@{\extracolsep{\stretch{0.5}}}*{7}{c}@{}}
 \hline
 Spaces & 1 & 2 & 3 & 4 & 5 & 6 & 7 \\
 \hline
 Room type & Conference & Conference & Conference & Conference & Private & Private & Private \\
 Number of position samples & 90  & 117& 121 & 169  & 48 & 56  & 128\\
 Number of valid position samples & 48  & 76 & 84& 111& 17 & 27 & 81  \\
 Percentage of valid position samples & $53.3\%$ & $65.0\%$ & $69.4\%$ & $65.7\%$ & $35.4\%$ & $48.2\%$ & $63.3\%$ \\
 \hline        
 \end{tabular*}
 \caption{Details of virtual space models. The number of position samples and percentage of valid position samples reflect the size of and how much the space is filled with objects. }
 \label{tab:spacedetails}
\end{table*}

\subsection{VR Experiments}
We implement a VR system simulating the telepresence of two users existing in their own spaces where each user's avatar is presented in the other user's space for interaction. For a clear explanation, we denote the local user and their avatar as X and X', and the remote user and their avatar as Y and Y'. Note that X will be a remote user from Y's point of view under our bi-directional telepresence. For experiments, X and X' are animated with upper-body tracking of the user while Y and Y' are static models with fixed placement and idle standing pose.

\subsubsection{Implementation}

\textit{\textbf{System.}} The application is implemented in Unity3D / SteamVR platform and uses the HTC Vive Pro set (HMD and two controllers) as a VR render and input device. The system consists of two identical hardware setups; each setup consists of a PC with an Intel i9 processor and an Nvidia Titan Xp graphics card. \\

\noindent \textit{\textbf{Virtual Character Control.}} The users manipulate the hand-held controllers to move their virtual characters in the virtual space. Viewpoint is fixed to the character head and only rotation tracking is allowed for the HMD. Hand transformations are obtained from controllers with respect to the tracked HMD, and the upper-body of virtual characters is animated by an inverse kinematics solver \cite{root2017final}. \\

\noindent \textit{\textbf{Spaces.}}
Considering our target scenario of the remote conferences in an MR telepresence environment, we carefully select seven realistic space models from 3Dwarehouse \cite{3dwarehouse} based on the variety of spatial composition and furniture arrangement. Figure \ref{fig:vrspaces} and Table \ref{tab:spacedetails} provide perspective-view images of selected models and their size and density, respectively.

\subsubsection{Experiment 1}
\label{sec:Exp1}
The proposed interaction feature and recommendation score are the basis for our visual guidance; we precede Experiment 1 to verify that the score of a local placement is consistent with the users' perception of the degree of interaction context preserved for the avatar placed at its OCP. To consider different levels of a discrepancy between two distant spaces, we select two pairs of conference-conference (Spaces 1 and 3) and conference-private (Spaces 4 and 7). \\

\noindent \textit{\textbf{Participants.}} We recruited $16$ participants ($6$ males / $10$ females) with the age range of $25$-$31$ ($\mu=26.5,\ \sigma=1.59$) in a local university community. All participants had experience with VR/AR applications before. \\

\noindent \textit{\textbf{Procedure.}}
We predefine three relations of interpersonal, human-object, and both; interpersonal and human-object relations correspond to X interacting with Y' and a screen in the local space, respectively.  
The range of recommendation scores to be presented to participants is chosen to be wide as $(0.5, 1.0]$, and it is divided into ten groups, separated by an interval of $0.05$.
A total of $60$ local - remote placement samples are generated with a combination of the space pairs, relations, and score groups. To generate a local-remote placement sample, we first randomly place Y and then place Y' at the OCP of Y placement, setting the screen as the interaction target. For X and X', we compute recommendation scores of all valid placements of X and randomly selected one within the target range.

Generated samples are presented to participants in a random order, and participants are asked to check the viewpoints of X (participant) and X' (their avatar) alternately while gazing and pointing at the given interaction targets. An image of three colored shapes (triangle, rectangle, and circle) is displayed on the screen as clear evaluation criteria, as shown in Figure \ref{fig:visoverview} a). After comparing the gaze and pointing context from viewpoints of X and X', participants answer a question, "My remote avatar placed at the remote space can transfer the contexts of my gaze and pointing accurately" on a 7-point Likert scale (1: Strongly Disagree - 7: Strongly Agree). \\

\begin{figure}[h]
\includegraphics[width=\linewidth]{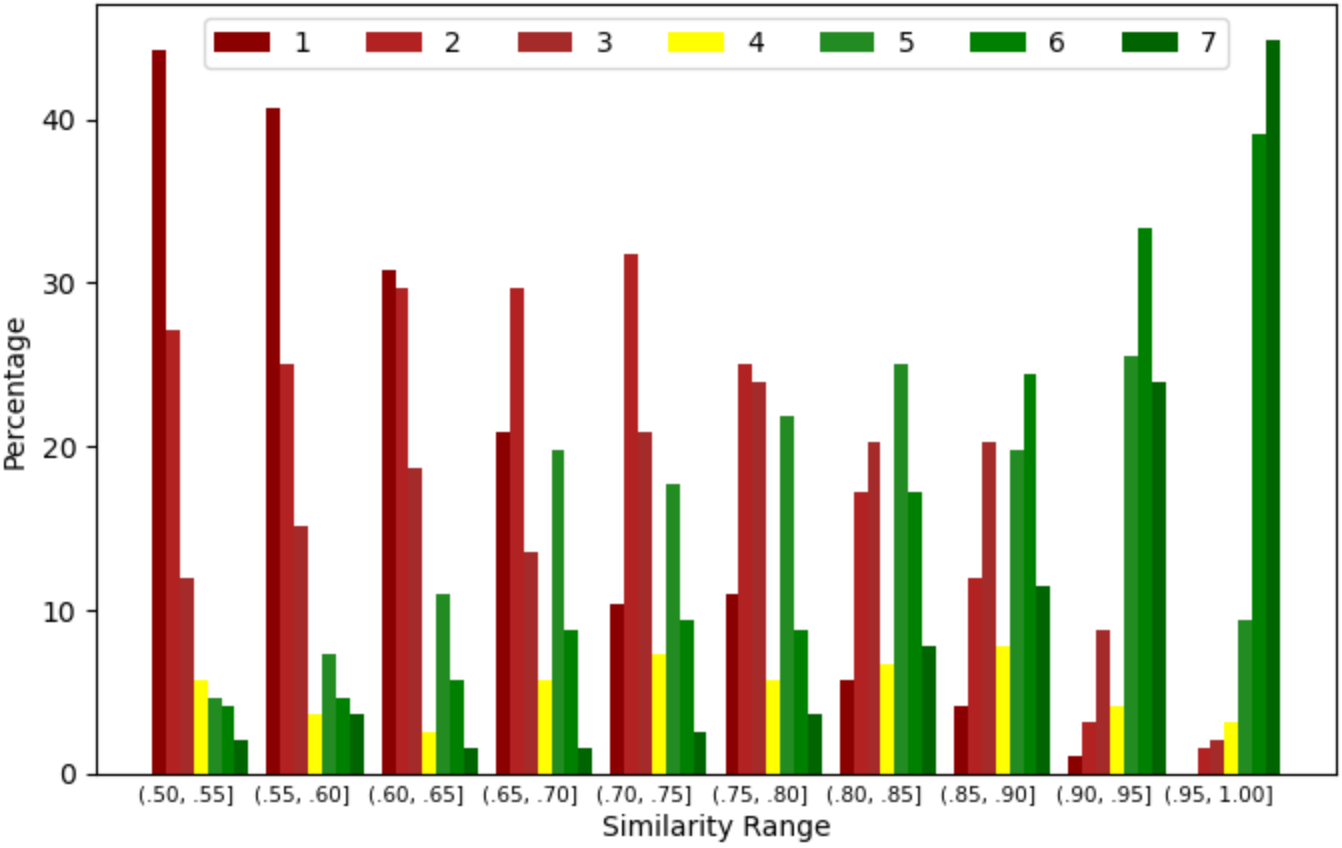}
\caption{The percentage of user responses on the degree of interaction context preserved for placement samples from score range groups in EXP1. The x-axis represents similarity groups where placements are sampled, and the y-axis represents the percentage of user responses. Each score group consists of seven bar graphs showing the percentage of score votes (ranging from 1 to 7) received by the sampled placements. The sum of the seven bar graphs is 100\%.}
\label{fig:responsesexp1}
\end{figure}

\noindent \textit{\textbf{Results and Analysis.}}
The correlation between the recommendation score and the user perception of gaze and pointing context preservation is examined by $\chi^2$ test. Figure \ref{fig:responsesexp1} and Table \ref{tab:chi2exp1} show the percentages of user responses by score groups and $\chi^2$ test result, respectively. A strong positive correlation is observed between the recommendation score (or the similarity of a placement pair) and the users' perception of the interaction context preservation (Spearman correlation$=0.602,\ p<.001$).

\begin{figure}[h]
    \centering
    \includegraphics[width=0.8\linewidth]{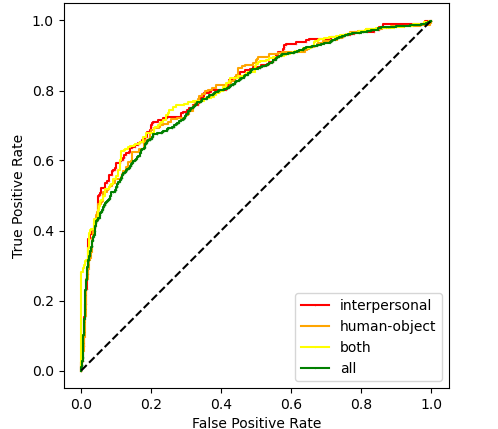}
    \caption{ROC curves of recommendation score for different relations in EXP1.}
    \label{fig:roc}
\end{figure}

To find the optimal threshold score that determines acceptable placement for interaction, we draw Receiver Operating Characteristic (ROC) curves of positive and negative placements; samples with user responses over scale 5 (Weakly Agree) are labeled as positive, or otherwise, labeled as negative. Figure \ref{fig:roc} shows the labeled sample distribution and the ROC curve. Table \ref{tab:rocexp1} shows the sensitivity and 1-specificity at various threshold values. Samples with interpersonal relations were found to have a higher threshold than the other relations, which presumably is because the user and the avatar have the same width and are less tolerable for divergence in viewpoints. According to the result of the entire samples, the best threshold value found is $.799$ and is rounded to $.8$ to be used for further experiments.

\begin{table}[h]
\centering
\begin{tabular}{lccc}
\hline
~ & Value & DoF & \textit{p}-value (2-sided)  \\
\hline
Pearson Chi-Square & $1003.817$ & $54$ & $.000$***  \\ 
Likelihood Ratio & $986.196$ & $54$ & $.000$***  \\ 
Linear-by-Linear Association & $688.240$ & $1$ & $.000$***  \\
Spearman Correlation & $.602$ & ~ & $.000$***  \\
\hline
\end{tabular}
\caption{$\chi^2$ test results of the user responses on degree of interaction context preserved across score groups in EXP1.}
\label{tab:chi2exp1}
\end{table}

\begin{table}[h]
    \centering
    \begin{tabular}{l c c c c c}
    \hline
    Relation & Area & \textit{p}-value & Thres. & Sens. & 1-Spec. \\
    \hline
    Interpersonal & .817 & .000 *** & .822 & .675 & .795 \\
    Human-Object & .809 & .000 *** & .807 & .688 & .801 \\
    Both & .817 & .000 *** & .799 & .628 & .886 \\
    \hline
    All & .801 & .000 *** & .799 & .675 & .795 \\
    \hline
    \end{tabular}
    \caption{ROC curve values of recommendation score for different relations in EXP1.}
    \label{tab:rocexp1}
\end{table}

\subsubsection{Experiment 2}
\begin{figure}[h]
\includegraphics[width=\linewidth]{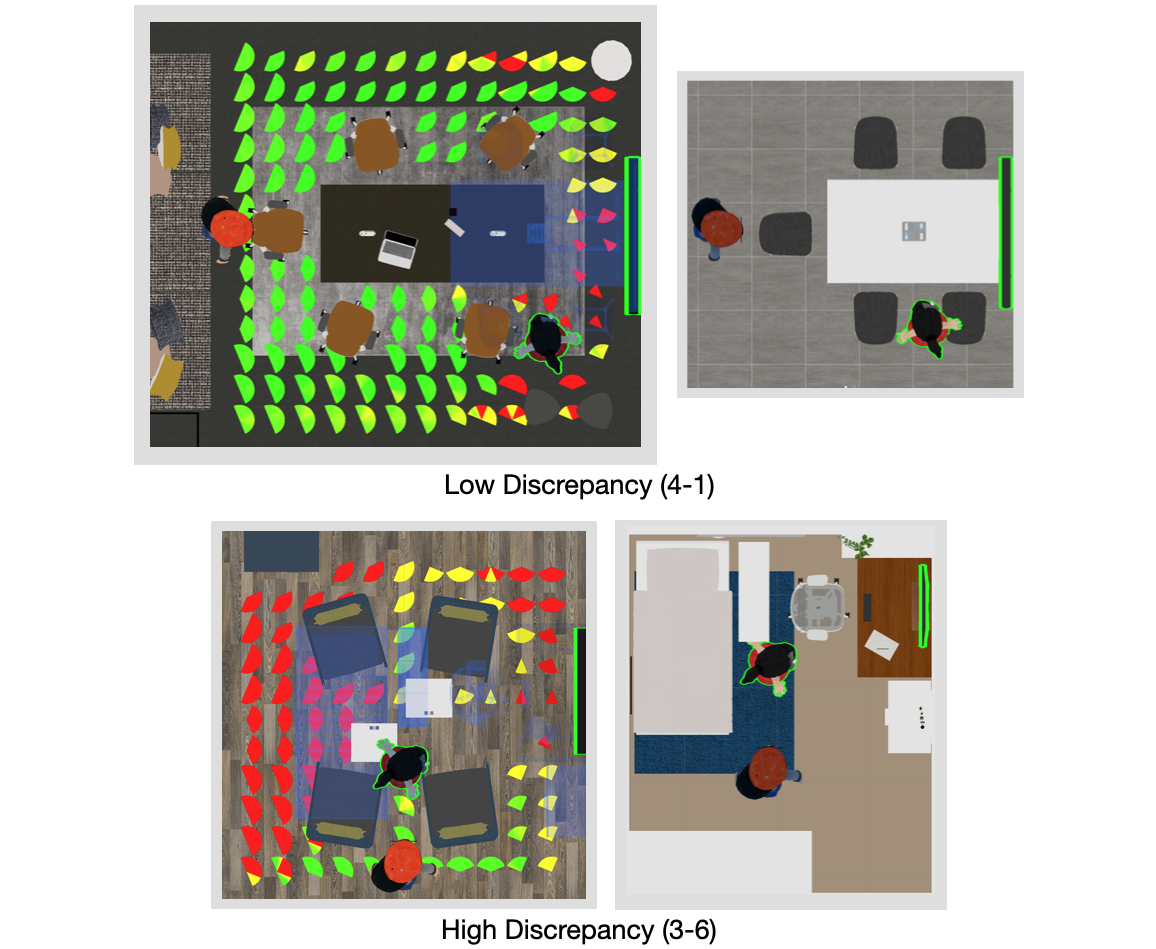}
\caption{Visual guidance examples for space pairs with different levels of discrepancy. Interaction targets are the partner's avatar and screen in both examples. For a similar pair (top), most of the placement samples are visualized as yellow-to-green sectors, meaning that the user has much freedom to choose good placements. For the dissimilar pair, the majority of the samples are represented with red sectors (bottom), informing that the interaction context can be preserved only in a narrow area.}
\label{fig:visdiscrepancy}
\end{figure}

The purpose of our visual guidance is to allow the users to choose the local placements where the interaction context can be well preserved by the synchronous motion of the remote avatar placed at the OCP. To measure the effect of visual guidance, we measure the recommendation scores of user-selected placements, while guided by different combinations of visualizations. To cover the diverse combinations of local-remote spaces with different levels of discrepancy in the layout, we select 4 pairs of 2 with the same functionality (conference-conference: 4-1, 3-2) and 2 with different functionalities (conference-private: 4-5, 3-6) from spaces shown in Figure \ref{fig:vrspaces}. Figure \ref{fig:visdiscrepancy} shows an example visual guidance of space pairs with low (4-1) and high (3-6) discrepancies. While the majority of local placements are visualized as valid (yellow to green color-coded sectors) for the pair with similar layouts, the number of invalid local placements (red sectors) increases with the spatial difference intensities. In addition, we measure the visual discomfort users felt from different combinations of visual guidance. \\

\noindent \textit{\textbf{Participants.}}  We recruited $16$ participants ($8$ males / $8$ females) with the age range of $25$-$31$ ($\mu=27.3,\ \sigma=2.25$) in a local university community. All participants had experience of VR/AR applications before and did not participate in Experiment 1. \\

\noindent \textit{\textbf{Procedure.}}
Similar to Experiment 1, we generate 64 problems from predefined variables of 4 space pairs, 4 relations (1 interpersonal, 1 human-object, and 2 both), and 4 combinations of visualizations. According to the amount of information given, we compare 4 combinations: Baseline (\textbf{B}) without any visual guidance, only showing invalid (red) sectors (\textbf{IV}), showing both invalid and valid sectors (\textbf{IV + V}), and finally transparent models added (\textbf{IV + V + Trans}). The placements for Y and Y' are sampled in the same way as in Experiment 1. Initially, the placement of X is randomly selected while X' is not visualized.

Before the experiment, we take a short warm-up session for participants to experience different visualizations and understand their meanings. Generated samples are provided in a random order, and participants are asked to move to a place of their choice to interact with given targets, guided by given visualizations; participants are not requested to follow the guidance. After a participant arrives at their destination, the OCP for their local placement is computed to place their avatar, and the score between X and X' placements is recorded for validation. At the end of the experiment, participants answer a question: "Visualizations provided cause visual overload." on a 7-point Likert scale (1: Strongly Disagree - 7: Strongly Agree). \\

\begin{figure}[h]
\centering
\includegraphics[width=0.95\linewidth]{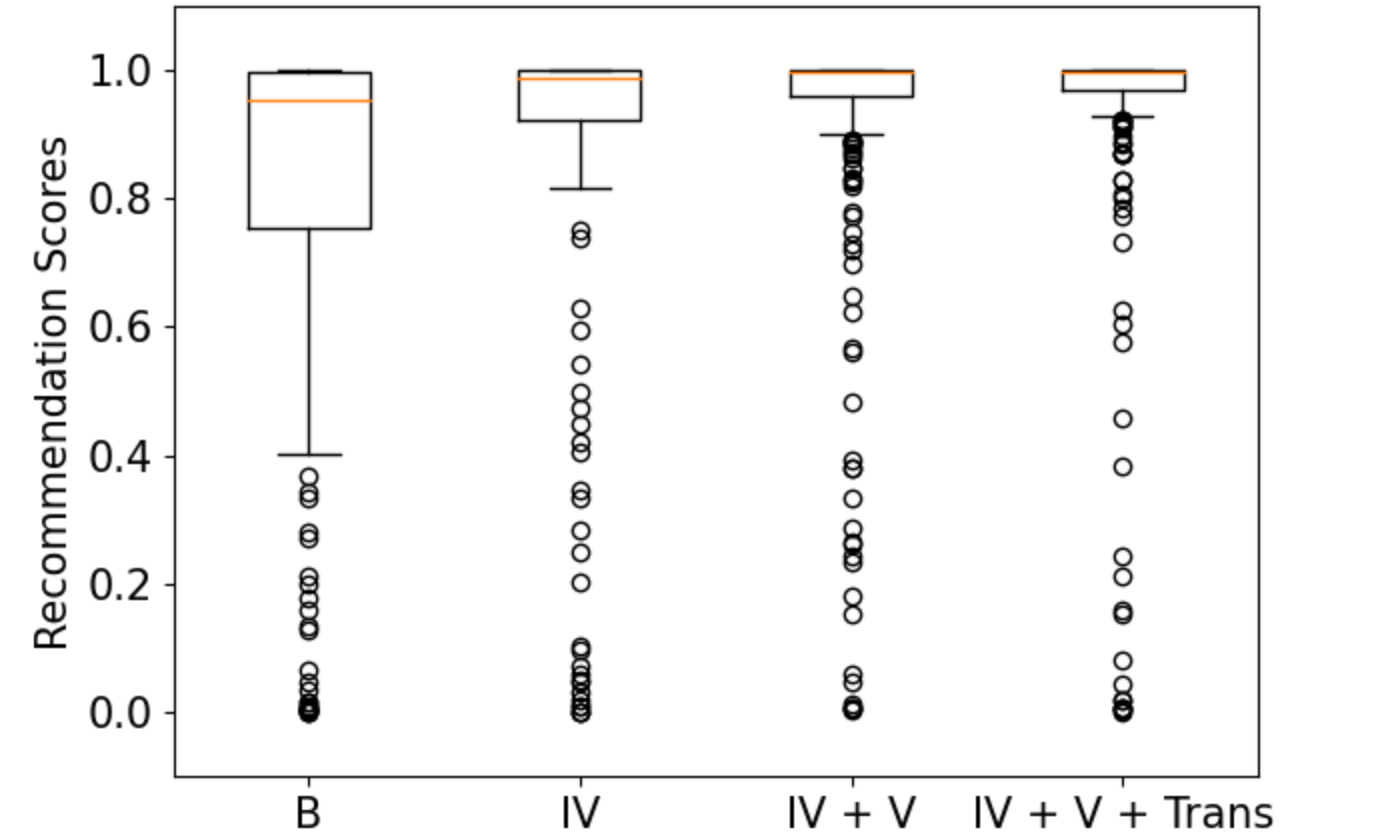}
\caption{Distribution of recommendation scores of user-selected placements for different combinations of visualizations in EXP2.}
\label{fig:userresponseExp2}
\end{figure}

\noindent \textit{\textbf{Results and Analysis.}}
Figure \ref{fig:userresponseExp2} shows the distribution of recommendation scores of the user placements from different combinations of visualizations. We perform the Kolmogorov-Smirnov test for the recorded scores for each visualization combination and observe that they are not normally distributed for all groups ($p<.001$). Therefore, we perform a Kruskal-Wallis test to examine the effect of different combinations of visualizations on interaction context by comparing the scores of the user-selected placements; we observe differences in the distribution of the recommendation scores among combinations of visualizations ($\chi^2 (1,\ N=1024)=63.84,\ p<.001$). The pairwise comparison results in Table \ref{tab:pairwisecomparison} show that adding visualizations of invalid and valid configurations has a significant effect on the scores of user placements ($p<.001$). However, adding transparent models of remote space objects is found not to significantly affect the scores of user-selected configuration ($p>.05$).

\begin{table}[h]
\centering
\begin{tabular}{llcc}
\hline
\multicolumn{2}{c}{Pair} & Stat. & \textit{p}-value \\
\hline
\textbf{B} & \textbf{IV} & $-85.88$ & $.001$ *** \\
\textbf{B} & \textbf{IV + V} & $-174.96$ & $.000$ *** \\
\textbf{B} & \textbf{IV + V + Trans} & $-180.04$ & $.000$ *** \\
\textbf{IV} & \textbf{IV + V} & $-89.08$ & $.001$ *** \\
\textbf{IV} & \textbf{IV + V + Trans} & $-194.16$ & $.000$ *** \\
\textbf{IV + V} & \textbf{IV + V + Trans} & $-5.08$ & $.846$ \\
\hline
\end{tabular}
\caption{Pairwise comparison result for combinations of visualizations in EXP2.}
\label{tab:pairwisecomparison}
\end{table}

\begin{figure}[h]
\centering
\includegraphics[width=0.95\linewidth]{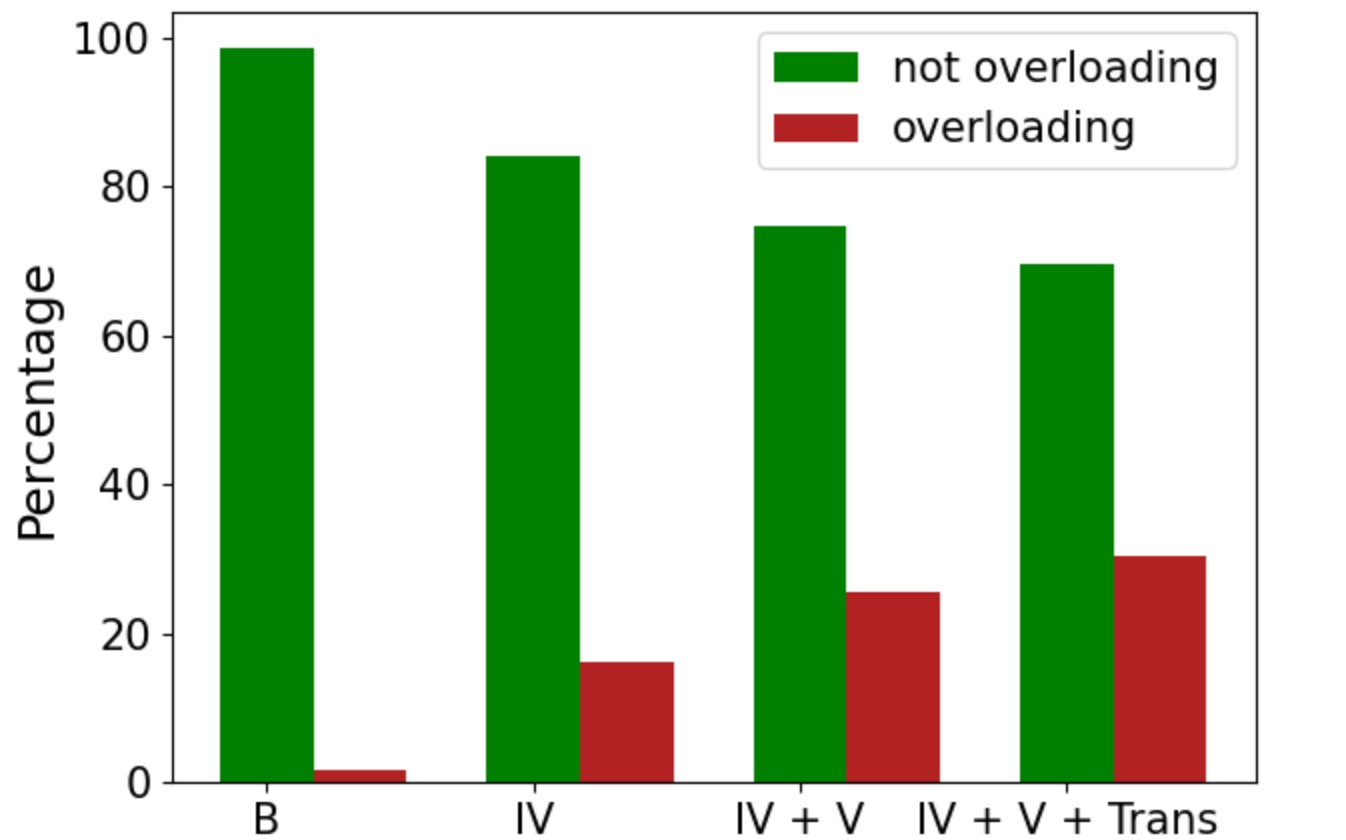}
\caption{User responses on visual overload for different combinations of visualizations in EXP2.}
\label{fig:userresponseExp2visover}
\end{figure}

To examine whether the proposed visual guidance causes discomfort for users, we code user responses over a scale of 5 as positive and others as negative for "causing visual overload." Figure \ref{fig:userresponseExp2visover} shows the percentages of positive and negative samples from different combinations of visualizations. Table \ref{tab:chi2exp2vis} shows the Chi-square test result, and a significant correlation was found between combinations of visualizations and visual overload ($\chi^2(1,\ N=1024)=82.115,\ p< .001$). Spearman correlation value of $.273$ states that the correlation is weak positive monotonic, which means that adding visualizations increases the visual overload users feel.

\iffalse
\begin{comment}
\begin{table}[h]
\centering
\begin{tabular}{lc}
\hline
Total Sample & $2048$ \\
Test Statistic & $114.82$ \\
DoF & $3$ \\
\textit{p}-value & $.000$ *** \\
\hline
\end{tabular}
\newline   
\caption{Kruskal-Wallis Test Summary}
\label{tab:kruskalwallis}
\end{table}
\end{comment}
\fi

\begin{table}[h]
\centering
\begin{tabular}{lccc}
\hline
~ & Value & DoF & \textit{p}-value (2-sided)  \\
\hline
Pearson Chi-Square & $82.115$ & $3$ & $.000$***  \\ 
Likelihood Ratio & $105.172$ & $3$ & $.000$***  \\ 
Linear-by-Linear Association & $78.779$ & $1$ & $.000$***  \\
Spearman Correlation & $0.278$ & ~ & $.000$***  \\
\hline
\end{tabular}
\caption{$\chi^2$ test results of user responses for visual overload across combinations of visualizations in EXP2.}
\label{tab:chi2exp2vis}
\end{table}

Based on these statistical results, we observed that including both invalid and valid placements increases the recommendation score for the user-selected local placement when interacting with given targets. The presence of transparent remote space objects overlaid in the local space did not significantly affect the recommendation scores. However, we have decided to include the entire set of visualizations (IV + V + Trans) as visual guidance for the MR experiment, simulating real use cases. This decision is based on our belief that including transparent remote space objects can provide explicit information about the layout of the remote space to the user. It is also worth mentioning that only 20\% of the participants reported experiencing visual overload.

\subsection{MR Experiment}\label{exp:Mrpilot}
To investigate the effect of the proposed method in the target telepresence scenario between dissimilar spaces, we design an MR user experiment in a remote presentation scenario. An example demo of the MR telepresence scenario is provided in the supplementary video. 

\subsubsection{Implementation}
\textit{\textbf{System.}} We implemented the MR system based on our VR system, and used ZED-mini and ZED cameras to display egocentric and room viewpoints, respectively. We employed the Photon Unity APIs for multiplayer network access and voice chat for verbal communication between the users. \\

\noindent \textit{\textbf{Virtual Characters.}} Different from the VR system with only upper-body tracking, we allow full-body tracking for the MR system to animate the remote avatar. To record the user's 2D position and orientation for computing similarity for visual guidance and avatar placement, we define the user reference frame by projecting the pelvis tracker's position and forward direction vector on the ground (XZ) plane; the up-vector is fixed to equal the world-y vector. The MR system does not require a virtual character model for the user. However, participants go through a calibration stage before the experiment, to adjust the scale and tracker orientation to animate the remote avatar model. \\

\begin{figure}[h]
\centering
\includegraphics[width=0.95\linewidth]{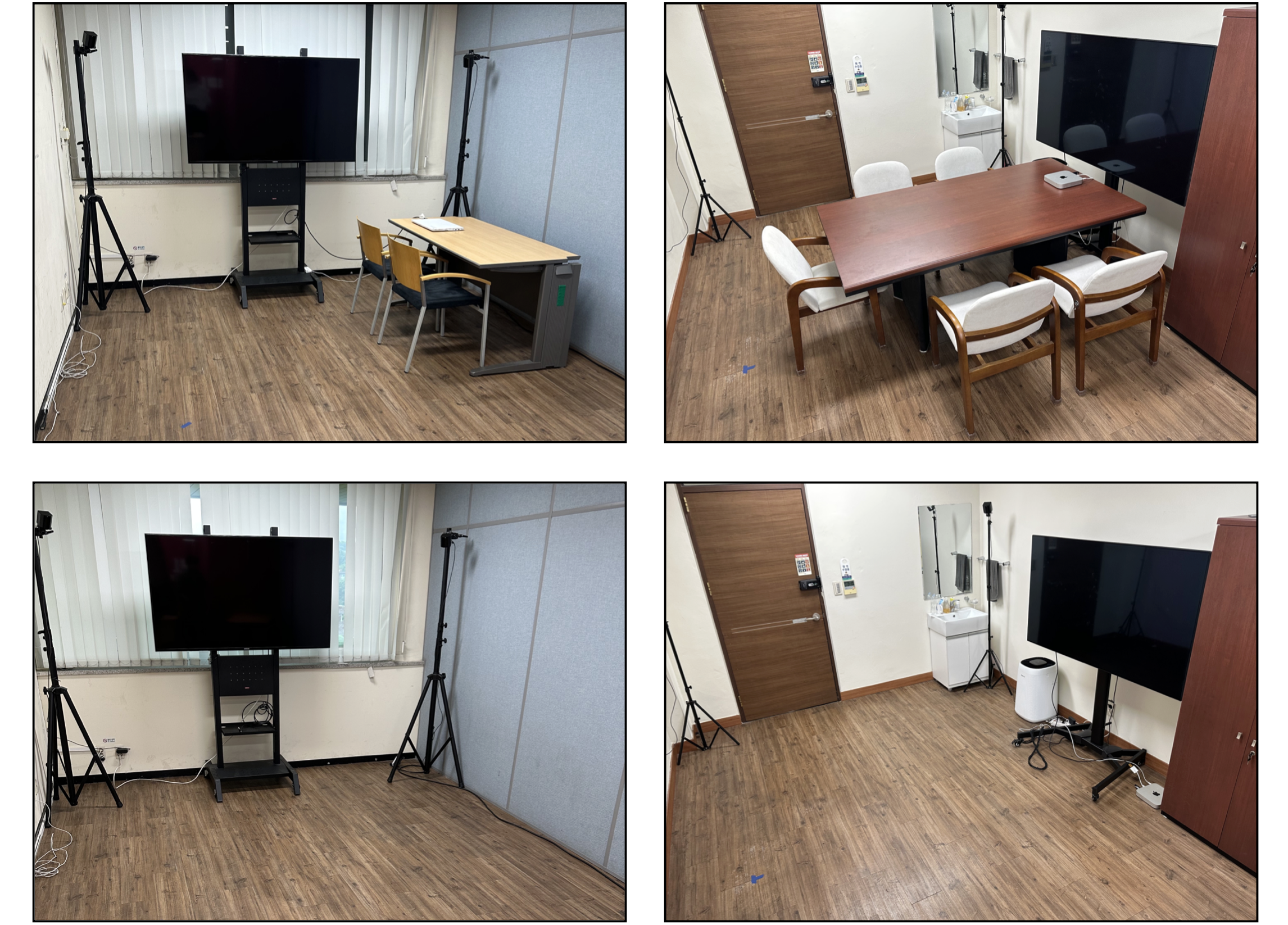}
\caption{Perspective view images of two pairs of remote spaces for the MR experiment. The upper row shows spaces O (left) and C (right), while the lower row displays their emptied versions, eO (left) and eC (right).}
\label{fig:mrspaces}
\end{figure}

\noindent \textit{\textbf{Spaces.}} Figure \ref{fig:mrspaces} shows perspective images of distant spaces O and C, and their emptied versions, eO and eC. Space O is designed as an office with open spaces in front of the screen, and space C has a typical conference room layout with a meeting table and multiple chairs. Identical virtual space models are manually constructed with precise scales of the objects. As part of the visual guidance, the virtual models of remote spaces are made transparent and overlaid onto the local spaces.

\subsubsection{Experiment}
The primary objective of the MR experiment is to validate the effectiveness of the proposed visual guidance and avatar placement methods in real user scenarios. To this end, we compared our method with the baseline method where neither visual guidance nor optimized avatar placement is given. In other words, participants will not see any visualization and be placed at the same location with respect to the interaction target in the condition baseline. In condition \emph{ours}, we enabled all visual guidance (IV + V + Trans) as well as optimized placement. To explore the impact of the level of spatial dissimilarity on user experience, we conduct the same experiment in space pairs of O-C and eO-eC on different days. As so in total, all participants experienced 2 conditions (baseline and ours) in each space pair. \\

\noindent \textit{\textbf{Participants.}} We recruited $12$ participants ($6$ male / $6$ females) with the age range between $23$ and $32$ ($\mu=28.25,\ \sigma=2.77$) in the local university community. All participants had MR experience. \\

\noindent \textit{\textbf{Procedure.}} The experiment is designed as a presentation session between two distant users. In this setup, one participant acts as the presenter in space O/eO and explains movie characters displayed on the screen, while the other participant in space C/eC serves as the listener. 
Each experiment session follows the following order:

\begin{enumerate}
  \item The listener selects the TV as the target and moves to a preferred location to listen to the presentation. After moving to the location, the listener places their avatar using controller input.
  \item The presenter selects the TV and the listener's avatar as the target and moves a preferred location to deliver the presentation. After moving to the location, the presenter places their avatar using controller input.
  \item The presenter initiates the presentation while the listener can freely interrupt with questions. The session lasts for a total of 5 minutes.
\end{enumerate}

Participants undergo two consecutive experiment sessions, starting with the baseline method followed by our proposed method. After each session, participants are asked to complete questionnaires. We chose to fix the order due to the potential bias, which we found during our initial pilot test with a smaller number of participants that they tend to remember information about the remote space, provided by visual guidance. In addition, we perform a post hoc semi-structured interview with all participants to further identify their preferences and limitations of the proposed visual guidance and placement method. \\

\noindent \textit{\textbf{Measures.}} After each session, participants answer the questions on a 7-point Likert scale (1: Strongly Disagree, 7: Strongly Agree), evaluating the remote user's avatar in my space (Q1-Q3) and visual guidance (Q4-Q6). The survey questions are as follows: \\

\begin{itemize}[leftmargin=*, after=]
    \item Q1: "The gaze and pointing performed by the remote user's avatar matched the context that the remote user verbally delivered."
    \item Q2: "The remote user's avatar did not make a collision with my environment." \cite{Kim2017Physicality}
    \item Q3: "I felt that the remote user's avatar  was aware of the objects in my space." \cite{Kim2017Physicality}
    
    \item Q4: "Visual guidance provided useful information to understand the placement of interaction targets in remote space."
    \item Q5: "Visual guidance helped me to select where to move for interaction." 
    \item Q6: "Visual guidance caused a visual overload." \\
\end{itemize}

\begin{figure}[h]
\centering
\begin{subfigure}
\centering
\includegraphics[width=0.95\linewidth]{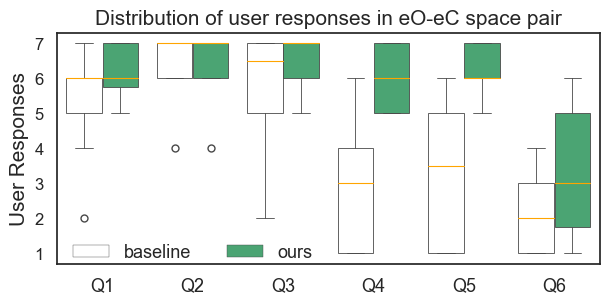}
\end{subfigure}
\begin{subfigure}
\centering
\includegraphics[width=0.95\linewidth]{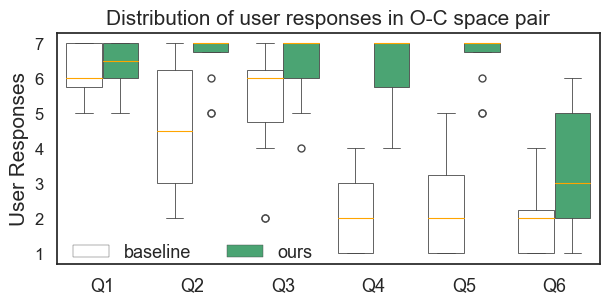}
\end{subfigure}
\caption{Distribution of user responses from the MR experiment in eO-eC space pair (top) and  O-C space pair (bottom).}
\label{fig:MRuserresponse}
\end{figure}

\noindent \textit{\textbf{Results, Analysis, and Discussion.}} Figure \ref{fig:MRuserresponse} presents distributions of user responses from the MR experiment. Since they did not pass the normality assumption, we used a non-parametric Wilcoxon signed rank test to compare two conditions: the baseline method (without visual guidance and identical placement) and ours (visual guidance and OCP) for two space pairs: eO-eC and O-C. The statistical results are presented in Table \ref{tab:mrexp}.

\begin{table}[h]
    \begin{minipage}{0.24\textwidth}
    \centering
    \captionsetup{skip=0pt}
    \caption*{eO-eC}
    \begin{tabular}{l c c c}
    \hline
    & W & z & \textit{p}-value \\
    \hline
    Q1 & 3.5 & -2.03 & .04* \\
    Q2 & 5 & 0 & 1 \\
    Q3 & 0 & -2.02 & .054 \\
    \hline
    Q4 & 0 & -3.06 & .002** \\
    Q5 & 0 & -2.93 & .004** \\
    Q6 & 5 & -2.07 & .042* \\
    \hline
    \end{tabular}
    \end{minipage}%
    \begin{minipage}{0.24\textwidth}
    \centering
    \captionsetup{skip=0pt}
    \caption*{O-C}
    \begin{tabular}{l c c c}
    \hline
    & W & z & \textit{p}-value \\
    \hline
    Q1 & 6 & -0.943 & .374 \\
    Q2 & 0 & -2.666 & 0.008** \\
    Q3 & 2.5 & -2.17 & .033* \\
    \hline
    Q4 & 0 & -3.059 & .002** \\
    Q5 & 0 & -3.059 & .002** \\
    Q6 & 9.5 & -1.835 & .072 \\
    \hline
    \end{tabular}
    \end{minipage}
    \caption{Statistical results of all questions between two conditions (baseline-ours) in our MR experiment.}
    \label{tab:mrexp}
\end{table}

Regarding avatar-related questions (Q1-Q3), we found weak significance ($p<.05$) for Q1 in the eO-eC space pair; participants rated slightly higher scores for ours (Baseline: M=5.33 SD=1.30, Ours: M=6.17 SD=0.83). We conjecture that our placement, preserving the angle-based features, resulted in a higher degree of context matching than the coordinates-preserving placement of the baseline, especially when the corresponding target objects have different sizes.

In the O-C space pair, Q2 ($p<.01$) and Q3 ($p<.05$) demonstrated significance, as avatars placed with the baseline method frequently penetrated remote space objects.

For visualization-related questions (Q4-Q6), we observed significance ($p<.01$) for Q4 and Q5 in both space pairs; participants assigned higher scores to ours (Q4 [eO-eC Baseline: M=2.92 SD=1.98, Ours: M=6.00 SD=0.86] [O-C Baseline: M=2.25 SD=1.14, Ours: M=6.33 SD=1.08] / Q5 [eO-eC Baseline: M=2.92 SD=2.02, Ours: M=6.33 SD=0.65] [O-C Baseline: M=2.42 SD=1.38, Ours: M=6.59 SD=0.80]), suggesting that they considered the visual guidance effective for understanding the remote space layout and selecting appropriate locations. For Q6, weak significance ($p<.05$) was observed in the eO-eC space pair. This may result from the emptiness of the spaces exacerbating the perception of visual overload.

At the end of the experiment, we asked participants three questions. The first question asked was to identify participants' direct opinions about whether they felt that visual guidance and avatar placement were effective. In condition eO-eC, 6 participants reported yes, 3 as ambiguous, and 3 as no while in condition O-C, all participants reported yes. The second question asked general advantages and disadvantages of the visual guidance they observed during the experiment. Third, their general opinion about MR collaboration communication. 

With statistical results and answered comments, we drive some discussion points. In the eO-eC pair, which resembles shared-space collaboration, about half of the participants found our method useful. Our visual guidance did not seem to affect the participants' choice of placement, as noted in the post-hoc interview. Participants mentioned that the overly broad green areas made it difficult to judge the effect of visual guidance, with comments like ``all spaces were colored as green". 

In the O-C pair, which assumes furnished environments for both local and remote spaces, all participants felt our method was better than the baseline. Many participants mentioned that visual guidance directly showed "where I should not go". Comments like: "I felt like I should go where I don't penetrate the objects of the remote space", and "It (Visual guidance) showed where I should be placed exactly" reveal the general reasons why they felt the method was useful. Unlike in the eO-eC pair, our visual guidance helped participants choose placements in the O-C pair, which appeared to impact their perception of necessity.

In eO-eC, Q6 showed weak significance ($p<0.05$) which implies that participants felt the visual guidance might cause visual overload. However, significance is not observed in O-C. This feeling of necessity seems to continue in the interview comment as well with words such as "I started to consider the remote space", and "I acknowledge where I should stand".
\section{Limitations}
Our experiments show that the proposed similarity measure is consistent with the user perception of context preservation, and visualizations, except transparent models, allow the users to select placements with higher similarities. Nevertheless, our work has limitations to be further explored.

\begin{figure}[h]
\centering
\includegraphics[width=0.9\linewidth]{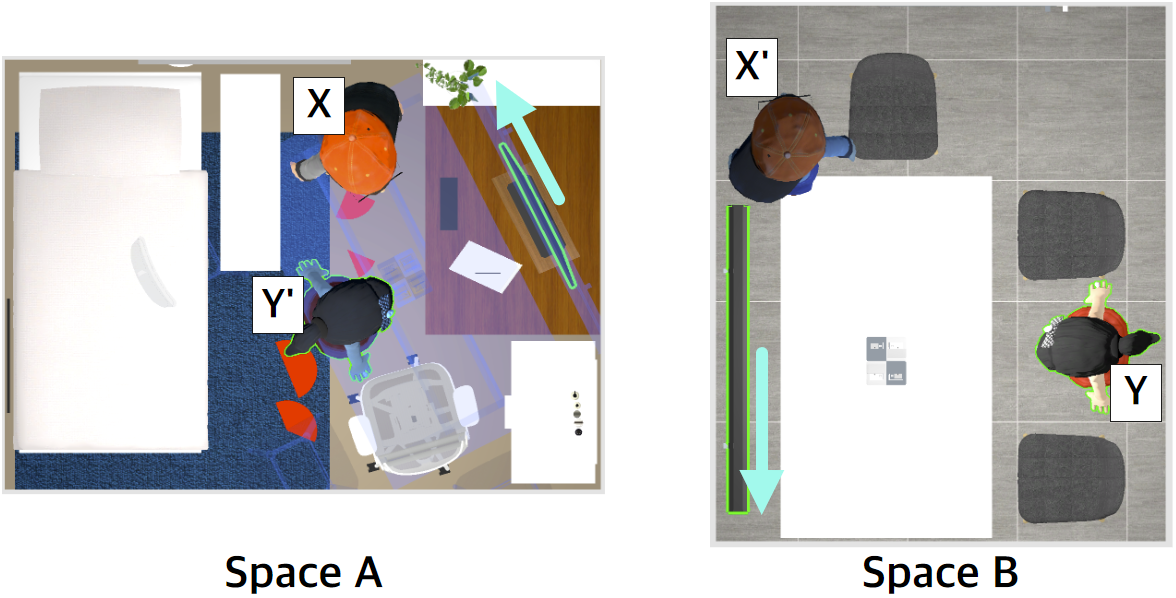}
\caption{A problematic case with limited free areas. In highly cluttered environments with small open areas, our visual guidance may struggle to identify feasible placements, resulting in the presentation of only red sectors. In the illustrated scenario, user X is positioned to the right of the display to engage with both the display and avatar Y'.
However, space B lacks an open area to the right of the display, causing avatar X' to be placed to the left of the display, facing away from the interaction targets. This renders telepresence interaction impossible. Cyan arrows denote the right direction of the displays.}
\label{fig:failclutter}
\end{figure}

Our method encounters challenges in spaces that are highly cluttered with limited open areas. Figure \ref{fig:failclutter} illustrates this problematic scenario. 
In space A, where user X wishes to interact with avatar Y' and the display, placement is constrained to the right side of the display because the left side is cluttered with a chair and a cabinet. On the other hand, in space B, empty areas are available only to the left of the display. In this situation, our visual guidance fails to provide green sectors indicating feasible placements. Consequently, when user X is positioned to the right of the display, avatar X' ends up placed to the left of the display, facing away from both the display and user Y, rendering telepresence interaction impossible.

Another limitation is that we only use angles in the 2D plane to describe the interaction feature between the user and targets, and may miss important information regarding the height of objects. For example, our method does not check the visual occlusion by tall objects between the user and target objects. Considering 3D scene information will enhance the generalizability of our approach to a wider variety of spatial layouts. Additionally, our out-of-space cost is designed for rectangular room shapes. Using differentiable signed distance functions can be employed to accommodate various room shapes.

As we narrowed our target scenario to remote meetings, space categories were limited to the conference and private rooms, and only gaze, and pointing interactions were considered. We need to validate and extend our method for other application scenarios, including interactive learning and remote collaboration with object manipulation. To allow touch interactions with objects, accessibility, and contact need to be maintained in the remote space. A possible approach would be considering the distance with interaction targets for avatar placement and using the head and arm gestures to transfer object manipulation context clearly.

Finally, our system assumes interaction between two users. As the number of users increases, the complexity of the problem can increase exponentially. It remains to tackle this high degree of complexity with many users and interaction targets.
\section{Conclusion}
In this paper, we presented a novel method of measuring and visualizing the suitability of a local placement for remote telepresence interaction. We modeled the angle-based interaction feature to represent the interaction and a Gaussian kernel similarity function to optimize remote placement. Our models have been experimentally confirmed to be coherent with user responses on the degree of interaction context preserved during the remote telepresence scenario. The proposed visualization methods showed a significant performance increase for remote interaction and successfully drove the users to the placements where the remote avatar placed at the OCP can preserve the user's interaction contexts.
\ifCLASSOPTIONcompsoc
  \section*{Acknowledgments}
\else
  \section*{Acknowledgment}
\fi
This work was supported by IITP, MSIT, Korea (2022-0-00566) and NRF, Korea (2022R1A4A5033689).

\ifCLASSOPTIONcaptionsoff
  \newpage
\fi
\bibliographystyle{IEEEtran}
\bibliography{refrences.bib}

%\input{9_appendix}
\vspace{10cm}

\begin{IEEEbiography}[{\includegraphics[width=1in,height=1.25in,clip,keepaspectratio]
{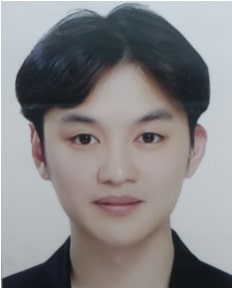}}]{Dongseok Yang} is a Ph.D. candidate with the Graduate School of Culture Technology at KAIST. He received an M.S. degree in Culture Technology from KAIST, Korea, in 2020 and a B.S. degree in Multimedia Engineering from Dongguk University, Korea, in 2018. His research interests include real-time human motion synthesis and MR telepresence.
\end{IEEEbiography}

\vskip -2\baselineskip plus -1fil
\begin{IEEEbiography}[{\includegraphics[width=1in,height=1.25in,clip,keepaspectratio]{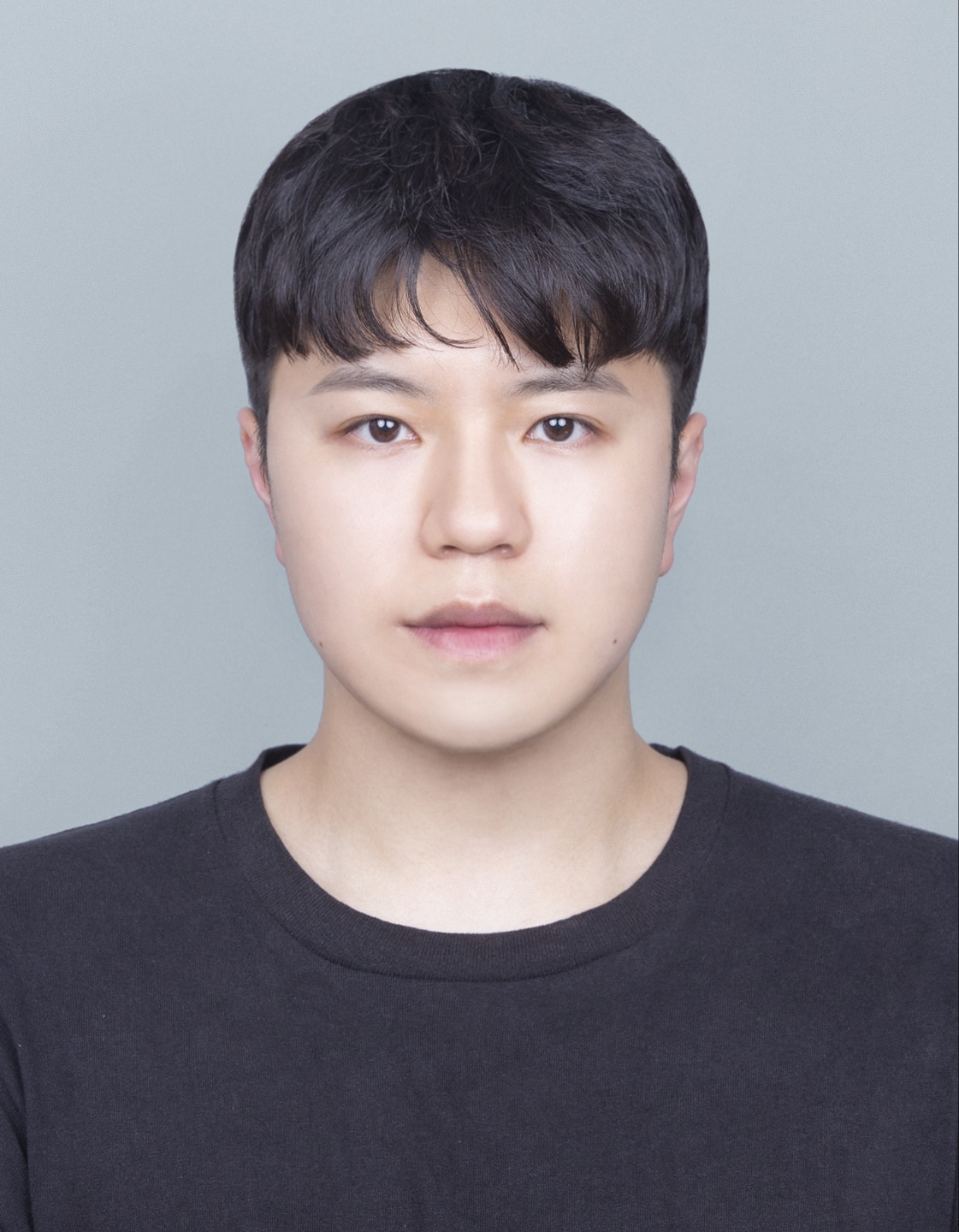}}]{Jiho Kang} is a Ph.D. candidate with the Graduate School of Culture Technology at KAIST. He received an M.S. degree in Culture Technology from KAIST, Korea, in 2023 and a B.S. degree in Electrical Engineering from Handong Global University, Korea, in 2021. His research interests include human animation and MR telepresence.
\end{IEEEbiography}

\vskip -2\baselineskip plus -1fil
\begin{IEEEbiography}[{\includegraphics[width=1in,height=1.25in,clip,keepaspectratio]{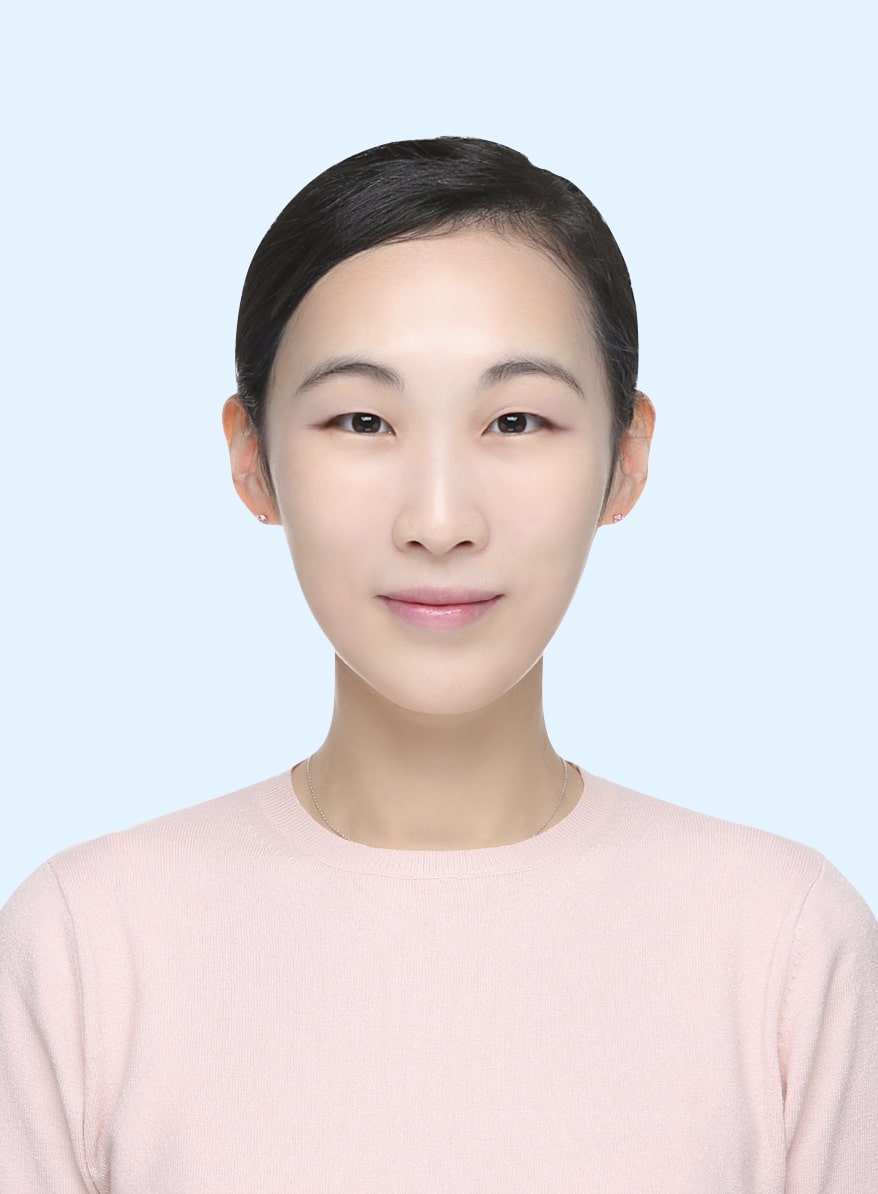}}]{Taehei Kim} is a Ph.D. candidate with the Graduate School of Culture Technology at KAIST. She received an M.S. degree in Culture Technology from KAIST, Korea, in 2021 and a B.S. degree in Asian Studies and Computer Science from Yonsei University, Korea, in 2018. Her research interests lie in the extent of adaptive human motion generation and perception in virtual reality.
\end{IEEEbiography}

\vskip -2\baselineskip plus -1fil
\begin{IEEEbiography}[{\includegraphics[width=1in,height=1.25in,clip,keepaspectratio]{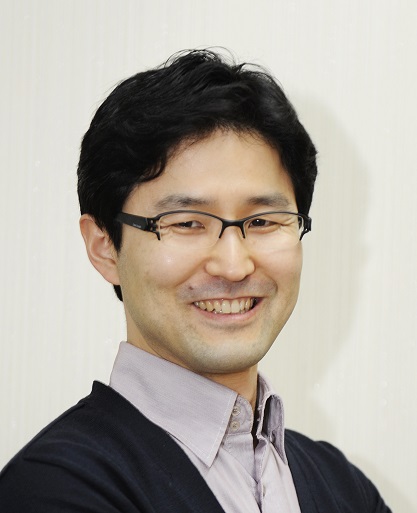}}]{Sung-Hee Lee} is a Professor with the Graduate School of Culture Technology at KAIST. His research interests include autonomous human animation, avatar motion generation, and human modeling. He received a Ph.D. degree in Computer Science from the University of California, Los Angeles, USA, in 2008, and a B.S. and M.S. degree in Mechanical Engineering from Seoul National University, Korea, in 1996 and 2000, respectively.
\end{IEEEbiography}

\end{document}